\documentclass[fleqn,usenatbib]{mnras}

\usepackage{newtxtext,newtxmath}
\usepackage[T1]{fontenc}
\usepackage{ae,aecompl}
\usepackage{rotating}
\usepackage{color}

\usepackage{graphicx}	% Including figure files
\usepackage[utf8]{inputenc}
\usepackage[dvipsnames]{xcolor}
\definecolor{royalfuchsia}{rgb}{0.79, 0.17, 0.57}
\title{The First Glimpse of ULXs Through the Near-Infrared Images Captured by the James Webb Space Telescope}

\author[S. Allak]{Sinan Allak\thanks{E-mail:0417allaksinan@gmail.com}
\\
Space Science and Solar Energy Research and Application Center (UZAYMER), University of Çukurova, 01330, Adana, Turkey\\
}

\date{Accepted 2023 October 27. Received 2023 October 23; in original form 2023 June 19}

\pubyear{2023}

\begin{document}
\label{firstpage}
\pagerange{\pageref{firstpage}--\pageref{lastpage}}
\maketitle

\begin{abstract}

This work presents the first results of near-infrared (NIR) counterparts of ultraluminous X-ray sources (ULXs) in the case of NGC 1672 by using \textit{James Webb Space Telescope} ({\it JWST}) observations. Through advanced astrometry, unique counterparts were identified for four of the eight ULXs (ULX-1, ULX-4, ULX5, and ULX-8) located in NGC 1672, while multiple counterparts were identified for the remaining ULXs. The NIR observations show clues of warm dust or circumbinary disk around ULX-5 and ULX-8. In the case of ULX-5, optical SED is a well-fitted blackbody with 4300 K while NIR excess is well-fitted by a blackbody with a temperature of 1600 K. The optical-NIR photometric results show that the possible donor star of ULX-5 may be a K-M type RSG (red supergiant) whose optical emission is affected by the presence of a circumbinary disk or warm dust. Additionally, the counterpart of ULX-4 could be an AGN (active galactic nuclei) or star cluster due to its high K-band magnitude. Thanks to the good enough resolution of the {\it JWST} images, in past studies, most of the point-like and/or bright NIR counterparts of ULXs in distant galaxies observed appear to be likely blended sources, so most likely, many of them do not have the red color that an RSG could have. The significant improvement in sensitivity and resolution supplied by {\it JWST} will lead to a new perspective on the ambiguous nature of ULXs donors and environments necessitating a significant reassessment of earlier infrared studies of ULXs.

\end{abstract}


\section{Introduction}           %% first-level sections will be auto-capitalized
\label{sect:intro}

Ultraluminous X-ray sources (ULXs) are non-nuclear point-like extragalactic sources with an isotropic X-ray luminosity (L$_{X}$ > 10$^{39}$ erg s$^{-1}$) exceeding the Eddington limit for a typical 10\(M_\odot\) stellar-mass black hole (sMBH) (see review by \citealp{2017ARA&A..55..303K,2021AstBu..76....6F} and \citealp{2023NewAR..9601672K}). Several hypotheses have been proposed to explain the nature of ULXs. In general, two main possibilities have been discussed. For example, many studies presented evidence of intermediate-mass black holes (IMBHs, 10$^{2}$- 10$^{5}$ M$\odot$) with accreting at sub-Eddington rates for the compact nature of ULX systems \citep{1999ApJ...519...89C,2004ApJ...614L.117M,2007Ap&SS.311..203R,2012MNRAS.423.1154S,2013MNRAS.436.3262C}. However, in current models, ULXs are associated with compact stellar remnants (neutron stars or black holes). For these models, geometric beaming and/or accretion at super-Eddington limits can explain the observed properties of ULXs \citealt{2007Ap&SS.311..203R,2009MNRAS.393L..41K,2018ApJ...857L...3W}.

The recent discovery of pulsations \citep{2014Natur.514..202B,2016ApJ...831L..14F,2017Sci...355..817I,2017MNRAS.466L..48I,2018MNRAS.476L..45C,2019MNRAS.488L..35S,2020ApJ...895...60R} as well as the detection of a cyclotron resonance scattering feature (CRSF) in the X-ray spectrum of M51 ULX-8 \citep{2018NatAs...2..312B,2019MNRAS.486....2M} and for the first Galactic PULX candidate Swift J0243.6+6124 \citep{2022ApJ...933L...3K} in a subset of ULX sources provides strong evidence in favor of at least some of these sources hosting neutron stars.

Optical studies offer valuable insights into the characteristics of the donor star, and disk geometry, and can help determine the mass of the accretor in ULX binaries. The observed optical emission in these systems can originate from the accretion disk, the donor star, or a combination of both. Many recent studies (\citealp{2012MNRAS.420.3599S,2014MNRAS.444.2415S,2018MNRAS.480.4918A,2019ApJ...884L...3Y,2022MNRAS.515.3632A,2022MNRAS.517.3495A}), focusing on optical variability, multi-band colors, and SED modeling, strongly suggest that the optical emission is likely contaminated or even dominated by reprocessed radiation from an irradiated accretion disk. As a specific example, the long-term optical light curves of the possible donors M51 ULX-4 \citep{2022MNRAS.510.4355A}, M51 ULX-8 \citep{2022MNRAS.517.3495A}, and NGC 1313 ULX-2 \citep{2009ApJ...690L..39L} have exhibited sinusoidal modulations. The best scenario reported for this optical modulation suggests that the accretion disk is the source of these modulations. However, it has been reported that the observed optical emissions of the ULX sources NGC 7793 P13 \citep{2014Natur.514..198M} and NGC 1313 X-2 \citep{2022MNRAS.511.5346S} come from the donors.

Multi-wavelength studies aiming to understand the properties of donor stars in ULXs have been conducted for over a decade. Moreover, the IR (infrared) wavelength observations, which are relatively less susceptible to the effects of extinction such as dust, offer valuable constraints for investigating the circumstellar/circumbinary environment of ULXs. Recently, the presence of warm dust with RSG donor star for NGC 300 X-1 has been reported by \citep{2019ApJ...883L..34H}. Therefore, multi-wavelength observations can be a good tool to avoid confounding multiple emitting components within ULXs (\citealp{2016ApJ...831...88D,2019ApJ...878...71L} and references therein). Some possible donor stars of ULXs which are bright in the near-infrared (NIR) bands might be red supergiants (RSGs) from IR spectroscopic and photometric analysis (\citealp{2020MNRAS.497..917L} and references therein). However, it should be emphasized that the identification of these ULXs was in general based on ground-based images or {\it Spitzer}\footnote{https://www.spitzer.caltech.edu/} can lead to false identifications in crowded fields due to large spatial resolutions.\\

This problem can be tackled with the very recent {\it James Webb Telescope} ({\it JWST}) observations. For example, the resolution of NIR images of the {\it NIRCam} camera is 0.03 and 0.06 arcsec/pixel which means 150 times better resolution than the best ground-based images for wide and medium filters, respectively. Therefore, this work is focused on NIR counterparts of ULXs in the case of NGC 1672, a late-type barred spiral galaxy at a distance of 16.3 Mpc \citep{2000AJ....119..612D}. This galaxy has been studied by \citealt{2011ApJ...734...33J}, who identified 9 ULXs, and \citep{2022MNRAS.515.3632A} who investigated possible optical counterparts of these ULXs. The primary goal of this work is to search for and identify possible NIR counterparts of these nine ULXs by using all available {\it JWST} images. The paper is structured as follows. Section \ref{sect:2} presents the properties of nine ULXs in galaxy NGC 1672, and this section also presents X-ray, {\it HST} optical, and {\it JWST} infrared observations of ULXs. Details of these observations and data reduction and analysis are presented in Section \ref{sec:3}. Section \ref{sec:4} presents results and discussions of the properties of counterparts. Finally, Section \ref{sec:5} summarizes the main conclusions of this study.\\

\section{Target sources \& Observations}
\label{sect:2}

\subsection{Ultraluminous X-ray Sources in NGC 1672}

X-ray study of the nearby barred spiral galaxy NGC 1672 was performed by \cite{2011ApJ...734...33J}. Nine out of 28 X-ray sources were identified as ULXs within the D$_{25}$ area of this galaxy. It is noteworthy that while source ULX-4 is located close to the center of the galaxy, the remaining ULXs are located in a spiral arm (see Figure \ref{F:rgb}). This distribution is not a surprise, since ULXs are often located in dense or star-forming regions of host galaxies. Of the nine ULXs, ULX-9 is not included in this study as it was not observed by both {\it HST} and {\it JSWT} instruments. Moreover, we, \cite{2022MNRAS.515.3632A} (hereafter {\it 2022b}), have investigated optical counterparts for these ULXs via improved astrometry. The location of the eight ULX examined in this study is given in Table \ref{T:astrometry}. Since this paper is also organized as a comparison of optical counterparts it is useful to remember the study of {\it 2022b}. The main results of the optical study are summarised below:\\

We have deployed archival data from {\it HST} optical and X-ray, to probe the nature of nine ULXs in NGC 1672. In order to determine possible donor stars, precise astrometry has been performed based on {\it Chandra} and {\it HST} observations. Unique optical counterparts for ULX-2 (X2) and ULX-6 (X6) and two counterparts for ULX-1, ULX-5, and ULX-7 have been determined within the astrometric error radius of 0.21 arcsec while no counterpart(s) have been identified for the remaining sources (ULX-3 (X3), ULX-4 (X4) and ULX-8 (X8)).\\

The photometric results of bright star groups in the environment of ULX-1 and ULX-7 suggest that their counterparts have similar properties to the nearby stars. Taking into account this result, we have estimated their ages as 20-30 Myr from the color magnitudes diagrams (CMDs). The absolute magnitudes of optical counterparts have been derived as $-$5 <M$_{V}$< $-$7.5 and with (B-A) spectral types.\\

The spectral energy distribution (SED) for the counterpart ULX-1 is represented with power-law with a photon index, $\alpha$$=$$-$ 2.13, and also shows 0.6 mag variability in UV-band. These are compatible with optical emissions arising primarily from an accretion disk. We have shown that the SED of ULX-2 is represented to a blackbody spectrum with B7$-$A3 type a supergiant donor star.\\ 

We have reported that the source ULX-6 is detected only in the I-band with 25.33 $\pm$ 0.11 mag. Additionally, We have calculated the X-ray variability factor as the ratio of maximum to minimum fluxes; this factor is found to be 5, 8, 30, 5, 50 for ULX-1, ULX-2, ULX-5, ULX-7, and ULX-9 respectively while for the remaining sources, the factor is $\leq$ 4. \\

\subsection{Observations of ULXs}

\begin{figure*}
\begin{center}
\includegraphics[angle=0,scale=0.3]{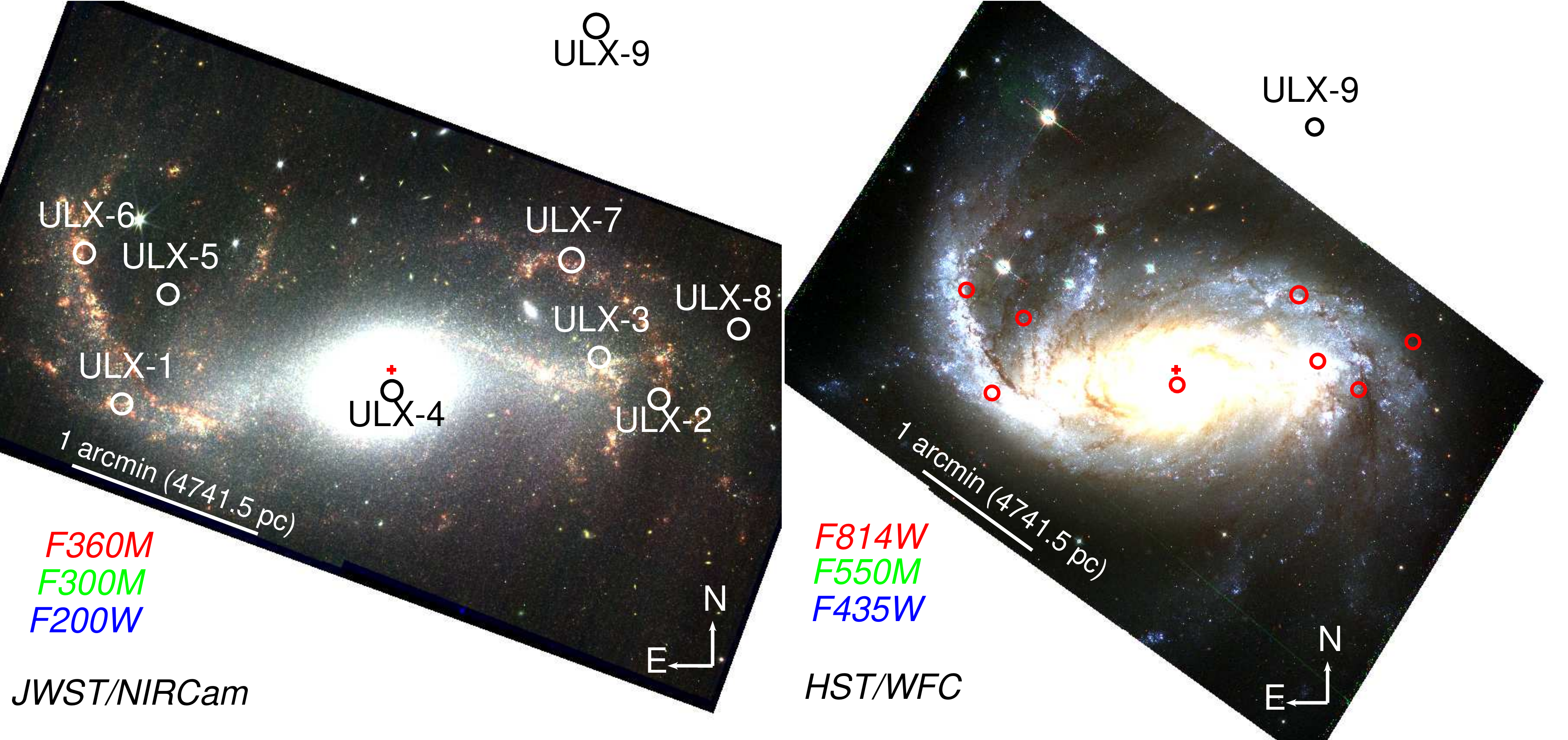}
\caption{The RGB images {\it JWST} (left) and {\it HST} (right) of the galaxy NGC 1672. The filters used for RGB images are highlighted on the images. The images are smoothed with a 3 arcsec Gaussian. The ULXs are indicated with white and red circles on the {\it JWST} and {\it HST}, respectively. For clarity, in the left panel, the position of the ULX-4 is indicated by the black circle. In both panels, the red cross indicates the center of the galaxy.}
\label{F:rgb}
\end{center}
\end{figure*}

The {\it JWST/NIRCam} observed the spiral galaxy NGC 1672 located at 16.3 Mpc on February 9, 2023, and the data set was published simultaneously. The galaxy {\it JWST} were captured with {\it NIRCam} F200W, F300M, F335M, and F360M and also {\it JWST/MIRI} F770W, F1000W, F1130W, and F2100W (proposal pi: Lee, Janice, proposal ID: 2107). ULXs are located in very crowded areas and it has been seen that sources that can be resolved in {\it NIRCam} images blend in {\it JWST/MIRI} images. NGC 1672 was also observed by {\it Hubble Space Telescope} ({\it HST} in 2005 and 2019 using the ACS/WFC (Advanced Camera for Surveys/Wide Field Channel) and WFC3/UVIS (The Wide Field Camera 3), respectively. In addition, these observations were used to compare counterparts of NIR with optical. NGC 1672 was also observed during 40 ks by {\it Chandra} ACIS (Advanced CCD Imaging Spectrometer) in 2006 (Obs ID: 5932). With this observation, the {\it GAIA}\footnote{https://www.cosmos.esa.int/web/gaia/dr2} source catalog is used for astrometry calculations in this study. The details of all the observations are given in Table \ref{T:obs}. RGB images {\it HST} and {\it JWST} of galaxy NGC 1672 are displayed in Figure \ref{F:rgb}.

\begin{table*}
\centering
\caption{The Log of Optical Near-infrared Observations}
\begin{tabular}{cccccccc}
\hline
Observatory/Instrument & Proposal ID & Date & Exp & Filter & Resolution & Resolution$^{*}$ & Pivot ($\lambda$)\\
& & (YYYY-MM-DD) & (s)& & (arcsec/pixel) & (pc/pixel) & ($\mu$m)\\
\hline
HST/ACS/WFC & 10354 & 2005-08-01 & 2444 & F438W & 0.050 & 4 pc & 0.433 \\
HST/ACS/WFC & 10354 & 2005-08-01 & 2444 & F550M & 0.050 & 4 pc & 0.558 \\
HST/ACS/WFC & 10354 & 2005-08-01 & 2444 & F814W & 0.050 & 4 pc & 0.805\\
HST/WFC3/UVIS & 15654 & 2019-06-24 & 2870 & F275W & 0.040 & 3.2 pc & 0.271\\
HST/WFC3/UVIS & 15654 & 2019-06-24 & 2480 & F336W & 0.040 & 3.2  pc & 0.335\\
HST/WFC3/UVIS & 15654 & 2019-06-24 & 1500 & F555W & 0.040 & 3.2  pc & 0.530\\
JWST/MIRI & 2107 & 2023-02-08 & 355.2 & F770W & 0.111 & 8.8 pc & 7.7\\
JWST/MIRI & 2107 & 2023-02-08  & 488.4 & F1000W & 0.111 & 8.8 pc & 10.0\\
JWST/MIRI & 2107 & 2023-02-08  & 1243.2 & F1130W & 0.111 & 8.8 pc & 11.3\\
JWST/MIRI & 2107 & 2023-02-08  & 1287.6 & F2100W & 0.111 & 8.8 pc &21.0\\
JWST/NIRCam & 2107 & 2023-02-09 & 2405 & F200W & 0.031 & 2.4 pc & 1.990\\
JWST/NIRCam & 2107 & 2023-02-09  & 773 & F300M & 0.063 & 5 pc & 2.996\\
JWST/NIRCam & 2107 & 2023-02-09  & 773 & F335M & 0.063 & 5 pc & 3.365\\
JWST/NIRCam & 2107 & 2023-02-09  & 858 & F360M & 0.063 & 5 pc & 3.621\\
\hline
\end{tabular}
\\ Note: $^{*}$ indicates the spatial resolution in parsec (pc) adopted at 16.3 Mpc distance of galaxy NGC 1672. \\
\label{T:obs}
\end{table*}

\section{Data Reduction and Analysis} \label{sec:3}

\subsection{Source Detection \& Photometry}

In order to identify positions and photometry for sources the {\scshape photutils v1.8}\footnote{https://photutils.readthedocs.io/en/stable/index.html\#} Astropy package for photometry was used. {\scshape photutils} primarily provides tools for source detecting and performing photometry of astronomical sources. For {\it JWST} data, background estimation, source detection, and photometry were performed by mostly following the steps outlined in \cite{2023ApJ...944L..26R}. Firstly, {\it detect\_threshold} and the {\it peak\_find} tasks were used to detect sources in all {\it NIRCam} images. The level of the background varies significantly across the image and therefore, the {\it photutils.background} task was used to estimate local background levels. Each source detection was more than 3-$\sigma$ above the local background. The aperture photometry was performed from a circular aperture with a radius of 3 pixels and the background was subtracted from an annulus nine pixels away from the source center. Vega magnitudes were obtained from flux densities in MJy/sr using the following Equation \ref{eq1}. Where {\it f1} is surface brightness in unit MJy/sr and {\it f2} is the flux of Vega in unit MJy and finally {\it PIXAR\_SR} is the average area of a pixel in sr (steradians), a constant provided in the image header keyword {\it PIXAR\_SR}.

\begin{equation}\label{eq1}
Vega_{mag} = -2.5 \times log(f1 \times PIXAR\_SR/f2)
\end{equation}

Moreover, for all drizzled images {\it HST} WFC and FWC3, point-like sources were detected with the {\it daofind} task and aperture photometry of these sources was performed using the {\it DAOPHOT} package \citep{1987PASP...99..191S} in {\scshape IRAF}\footnote{https://iraf-community.github.io/} (Image Reduction and Analysis Facility). To perform aperture photometry, 3 pixels (0$\arcsec$.15) aperture radius and for the background, nine pixels were chosen. In order to obtain photometric errors and data quality propagation, the pixel values were multiplied by the exposure time using {\it imarith} tool in {\scshape iraf}. The Vega magnitudes were derived instrumental magnitudes using ACS/WFC zero point magnitudes (ZPM) taken from the ACS Zeropoints Calculator\footnote{https://acszeropoints.stsci.edu/} and WFC3/UVIS ZPM were taken from the study \cite{2022AJ....164...32C}. To derive the aperture corrections a similar approach of \citealp{2022MNRAS.517.3495A} was followed.

\subsection{Determination of Counterparts}

\begin{figure*}
\begin{center}
\includegraphics[angle=0,scale=0.4]{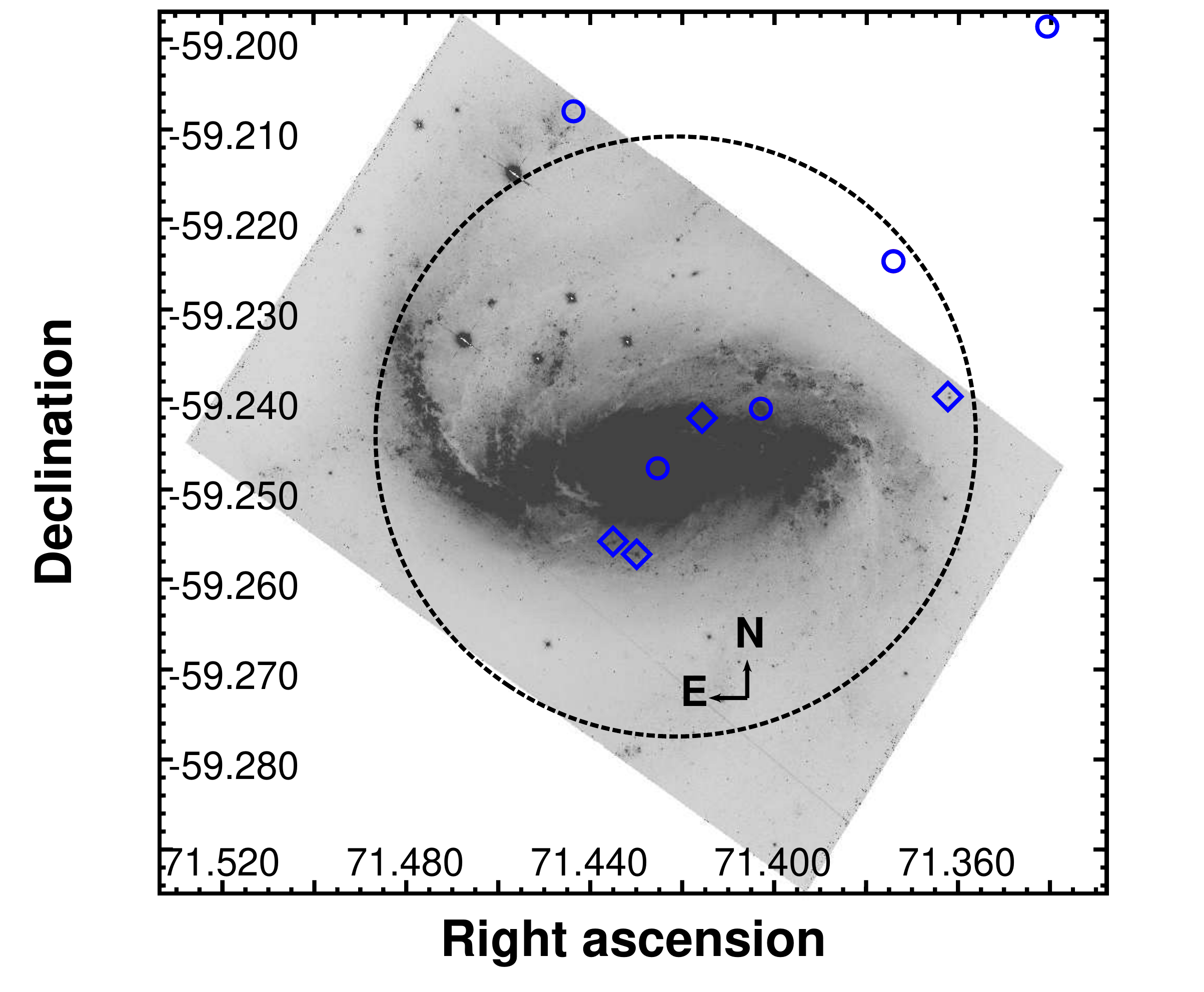}
\caption{Positions of reference sources between {\it Chandra} and {\it GAIA} (solid blue circles) and between {\it GAIA} and {\it JWST} (solid blue diamonds) are shown on the {\it HST/WFC} F550M image used in astrometric calculations for ULXs in NGC 1672. The black dashed circle represents the moderate offset to the optical axis for the {\it Chandra} image with radius of 2.5 arcmin.}
\label{F:astroJ}
\end{center}
\end{figure*}

Precise astrometry is necessary to determine both the NIR and optical counterparts of the eight ULXs, whose positions are provided in Table \ref{T:astrometry} and are shown in Figure \ref{F:rgb} on the {\it HST} and {\it JWST} RGB images. NIR counterparts of ULXs can be determined by precise astrometric calculations using {\it NIRCam} and {\it Chandra} observations, which generally have a very good spatial resolution. The reference sources were searched by comparing {\it Chandra} X-ray image with {\it NIRCam} observations. 58 X-ray sources were detected in {\it Chandra} image by using {\it wavdetect} tool in {\it Chandra} Interactive Analysis of Observations ({\scshape ciao})\footnote{https://cxc.cfa.harvard.edu/ciao/}. Only 22 out of 58 X-ray sources matched the {\it NIRCam} images. The 22 X-ray sources were compared with NIR point sources. However, since adequate reference sources were not found for astrometric calculations between {\it Chandra} and {\it NIRCam} images, Gaia Data Data Release 3 {\it GAIA/DR3}\footnote{https://www.cosmos.esa.int/web/gaia/data-release-3} source catalog was used for astrometric calculations. This catalog searched the reference sources that appear to be isolated and point-like in {\it Chandra} and {\it NIRCam} images. Taking account of the same shift directions, except the core of the galaxy four reference sources were found for both between both {\it Chandra}-{\it GAIA} and {\it GAIA}-{\it JWST} (F200W) (see Figure \ref{F:astroJ}). Almost all of these references are located in the moderate offset to the {\it Chandra} optical axis with a radius of 2.5 arcmin.\\

The astrometric offsets between {\it Chandra} and {\it GAIA} were found as -0$\arcsec$.17 $\pm$ 0$\arcsec$.17 for R.A and 0$\arcsec$.02 $\pm$ 0$\arcsec$.10 for Decl. with 1-$\sigma$ errors and also astrometric offsets between {\it GAIA} and {\it JWST} were found as 0$\arcsec$.03 $\pm$ 0$\arcsec$.03 for R.A and -0$\arcsec$.03 $\pm$ 0$\arcsec$.01 for Decl. with 1-$\sigma$ errors. As seen in Figure \ref{F:astroJ}, especially one out of four references between {\it Chandra} and {\it GAIA} has a large offset to {\it Chandra} the optical axis ($\sim$ 3.6 arcmin) but, it does not affect the astrometric error radius, since the shift and shift direction are the same as the other references. The total astrometric errors between {\it Chandra}-{\it GAIA} and between {\it GAIA} - {\it JWST} were derived as 0$\arcsec$.20 and 0$\arcsec$.03, respectively. Following the study of \cite{2022MNRAS.517.3495A}, the positional error radius was derived as 0.38 arcsec at a 90 percent confidence level that this radius corresponds to 30 parsecs (pc) adopted the distance of galaxy NGC 1672 (79 pc/arcsec). Moreover, relative astrometry was performed between {\it NIRCam} and {\it HST} images to determine the optical counterparts of NIR counterparts using {\it GAIA/DR3} source catalog. The total astrometric errors between {\it JWST/NIRCam} - {\it HST/WFC} and between {\it JWST/NIRCam} - {\it HST/WFC3} were derived as 0$\arcsec$.14 and 0$\arcsec$.1, respectively at 90\% confidence level.\\

Since the ULXs are located in the crowded star fields, even the smallest shifts between filters will change the number of counterparts and may mean that we examine the wrong possible donor star. For this reason, shift values between filters were checked for each instrument. Accordingly, 0.01 and 0.06 arcsec shift values were calculated for R.A and for Decl between {\it NIRCam} wide (F200W) filter and medium filters (F300M, F335M, and F360M). Between {\it NIRCam} and {\it MIRI} filters, shifts of 0.16 and 0.03 arcsec were determined for R.A and Decl, respectively. In the case of filters of {\it HST/WFC} and {\it HST/WFC3}, there were not found any shifts. However, shifts of 0.2 and 0.3 arcsec were determined for R.A and Decl between {\it HST/WFC} and {\it HST/WFC3} filters, respectively. Counterparts were searched taking into account the determined shifts.

\begin{figure*}
\begin{center}
\includegraphics[angle=0,scale=0.35]{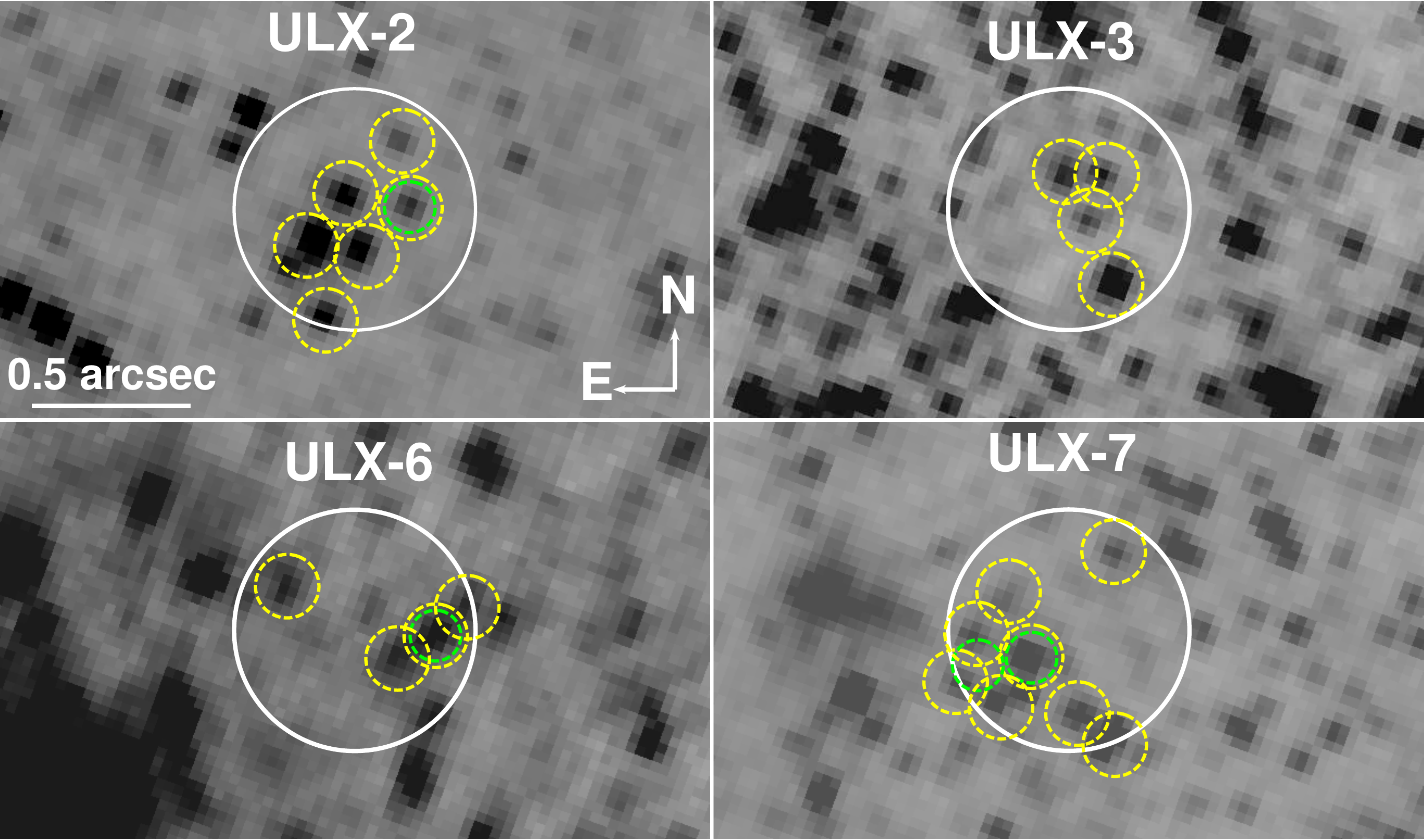}
\caption{{\it NIRCam} F200W image of corrected X-ray positions of ULX-2, ULX-3, ULX-6, and ULX-7 which have multiple NIR counterparts. The dashed yellow circles (0.1 arcsec) represent NIR counterparts detected with above 3-$\sigma$ threshold and the white circles represent an astrometric radius of 0.38 arcsec. The dashed green circles (0.08 arcsec) represent the optical counterparts determined by {\it 2022b}. All panels are the same scale and the north is up and the east is left.}
\label{F:multi}
\end{center}
\end{figure*}

\begin{figure*}
\begin{center}
\includegraphics[angle=0,scale=0.3]{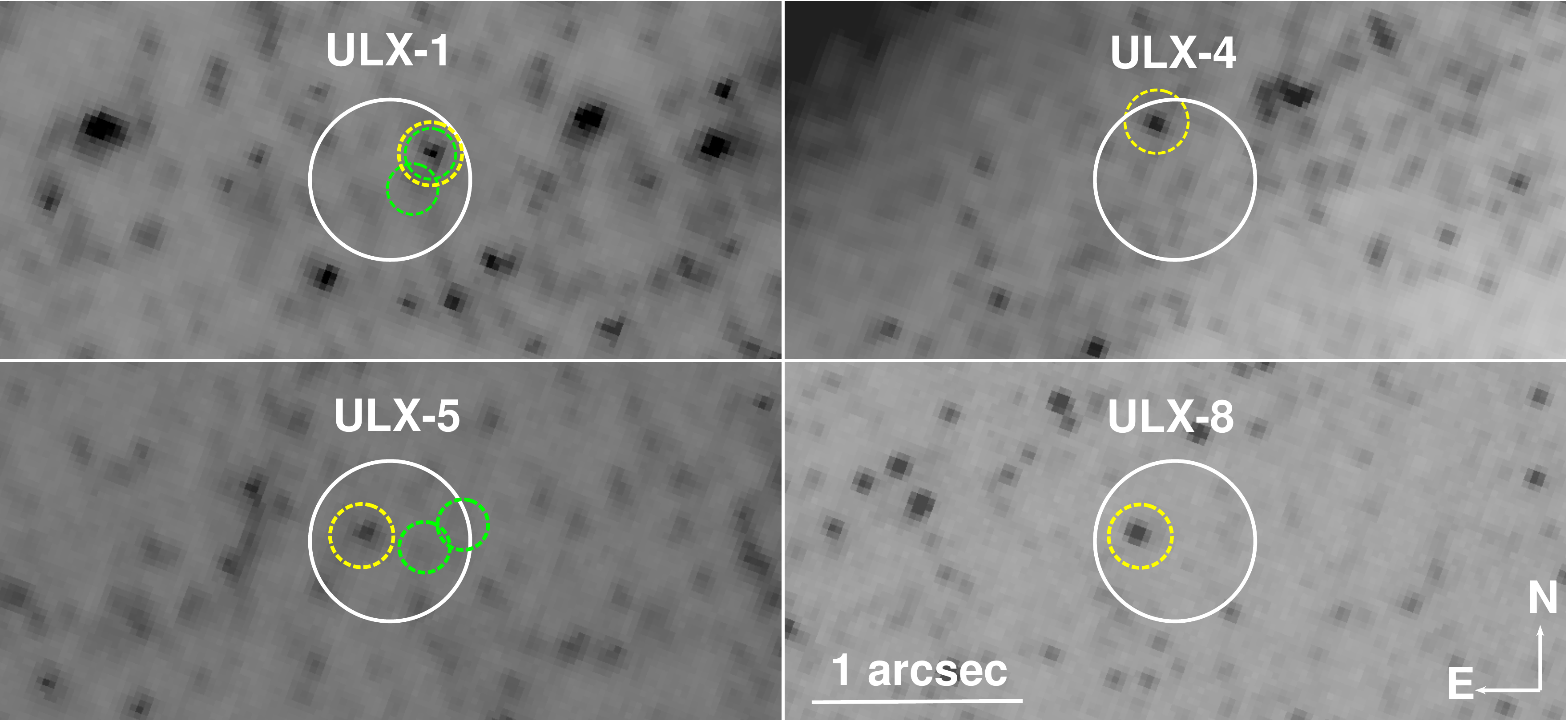}
\caption{{\it NIRCam} F200W image of corrected X-ray positions of ULX-1, ULX-4, ULX-5, and ULX-8 which have unique NIR counterparts. Other explanations are the same as in Figure \ref{F:multi}.}
\label{F:uniq}
\end{center}
\end{figure*}

Table \ref{T:astrometry} shows the corrected X-ray coordinates of the eight ULXs for NIR and optical images. The multiple NIR counterparts of ULXs are, marked on the {\it NIRCam} F200W image, displayed in Figure \ref{F:multi} and for unique NIR counterparts displayed in Figure \ref{F:uniq}. Moreover, the optical counterparts identified in our previous study are also shown in these Figures with dashed green circles. The derived Vega magnitudes of the optical and NIR counterparts are given in Table \ref{T:fotometri}. The identifications of the NIR and optical counterparts for each of the ULX were presented in detail following:\\ 

\begin{table*}
\centering
\caption{Coordinates of the Chandra X-ray and {\it NIRCam} NIR counterparts of ULXs.}
\begin{tabular}{cccrccccccc}
\hline
Source number & {\it Chandra} R.A.& {\it Chandra} Decl.& {\it JWST} R.A.& {\it JWST} Decl. \\
... & (hh:mm:ss.sss) & ($\degr$ : $\arcmin$ : $\arcsec$)  & (hh:mm:ss.sss) & ($\degr$ : $\arcmin$ : $\arcsec$) \\
\hline
ULX-1 & 4:45:52.823 & -59:14:56.15  & 4:45:52.854 & -59:14:56.20  \\
ULX-2 & 4:45:31.603 & -59:14:54.68  & 4:45:31.632 & -59:14:54.70  \\
ULX-3 & 4:45:33.965 & -59:14:42.03  & 4:45:33.995 & -59:14:42.06  \\
ULX-4 & 4:45:42.166 & -59:14:52.17 & 4:45:42.181 & -59:14:52.23  \\
ULX-5 & 4:45:50.996 & -59:14:22.96  & 4:45:51.024 & -59:14:22.96  \\
ULX-6 & 4:45:54.289 & -59:14:10.44 & 4:45:54.325 & -59:14:10.47  \\
ULX-7 & 4:45:35.078 & -59:14:12.68 & 4:45:35.089 & -59:14:12.63  \\
ULX-8 & 4:45:28.453 & -59:14:33.39 & 4:45:28.489 & -59:14:33.46  \\
\hline
\label{T:astrometry}
\end{tabular}
\end{table*}

\begin{table*}
\centering
\caption{The derived reddened Vega magnitudes of the possible optical and NIR counterparts.}
\begin{tabular}{cccccccccc}
\hline
 Counterparts & UVIS/F336W & UVIS/F555W & ACS/F435W & ACS/F550M  & ACS/F814W \\
{\it ULX-5} & 28.1 & 26.47 $\pm$ 0.13 &  26.57 $\pm$ 0.22 & 26.35 $\pm$ 0.17 & 24.81 $\pm$ 0.06 \\
\hline
 & NIRCAM/F200W & NIRCAM/F300M & NIRCAM/F335M & NIRCAM/F360M  & ACS/F814W \\
\hline
{\it ULX-1} & 21.71 $\pm$ 0.06 & 22.57 $\pm$ 0.06 & 20.43 $\pm$ 0.05 & 21.01 $\pm$ 0.04 & 24.22 $\pm$ 0.06\\
{\it ULX-4} & 18.66 $\pm$ 0.04 & 19.99 $\pm$ 0.03 & 18.71 $\pm$ 0.03 & 20.35 $\pm$ 0.03 & ... \\
{\it ULX-5} & 22.59 $\pm$ 0.04 & 22.42 $\pm$ 0.03 & 22.49 $\pm$ 0.03 & 22.34 $\pm$ 0.04 & ... \\
{\it ULX-8} & 22.57 $\pm$ 0.04 & 22.63 $\pm$ 0.03 & 22.92 $\pm$ 0.03 & 22.97 $\pm$ 0.03 & ... \\
\hline
\end{tabular}
\label{T:fotometri}
\end{table*}

{\it ULX-1}: By examining all the NIR images listed in Table \ref{T:obs}, a unique NIR counterpart for ULX-1 was identified within the astrometric error radius (see Figure \ref{F:uniq}). In the case optical study ({\it 2022b}), two optical candidates ({\it X1$\_1$} and {\it X1$\_2$}) have been determined within the respective error radius of 0.21 arcsec (see green dashed circles in Figure \ref{F:uniq}). This NIR counterpart has matched the source {\it X1$\_2$} which has been detected by only red filter of {\it HST/WFC3} F814W. As we have reported in {\it 2022b}, the source {\it X1$\_1$} is more dominant in the UV band, but in this study, it could not be detected at the 3-$\sigma$ threshold in {\it JWST} images. In this case, ULX-1 has two possible donors, one is dominant in the UV-optical bands and the other is dominant in the NIR bands. \\

{\it ULX-2}: Six NIR counterparts were identified for ULX-2 within the astrometric error radius at 3-$\sigma$ detection threshold, one of which corresponds to the previously reported optical counterpart (see green dashed circle in Figure \ref{F:multi}). \\

{\it ULX-3}: The four possible NIR counterparts were determined within the error radius while in the case of the optical wavelengths, any optical source(s) has not been determined.\\

{\it ULX-4}: A unique NIR counterpart was determined within the error radius. Just as with the source ULX-3, the optical counterpart(s) were not determined.\\

\begin{figure*}
\begin{center}
\includegraphics[angle=0,scale=0.3]{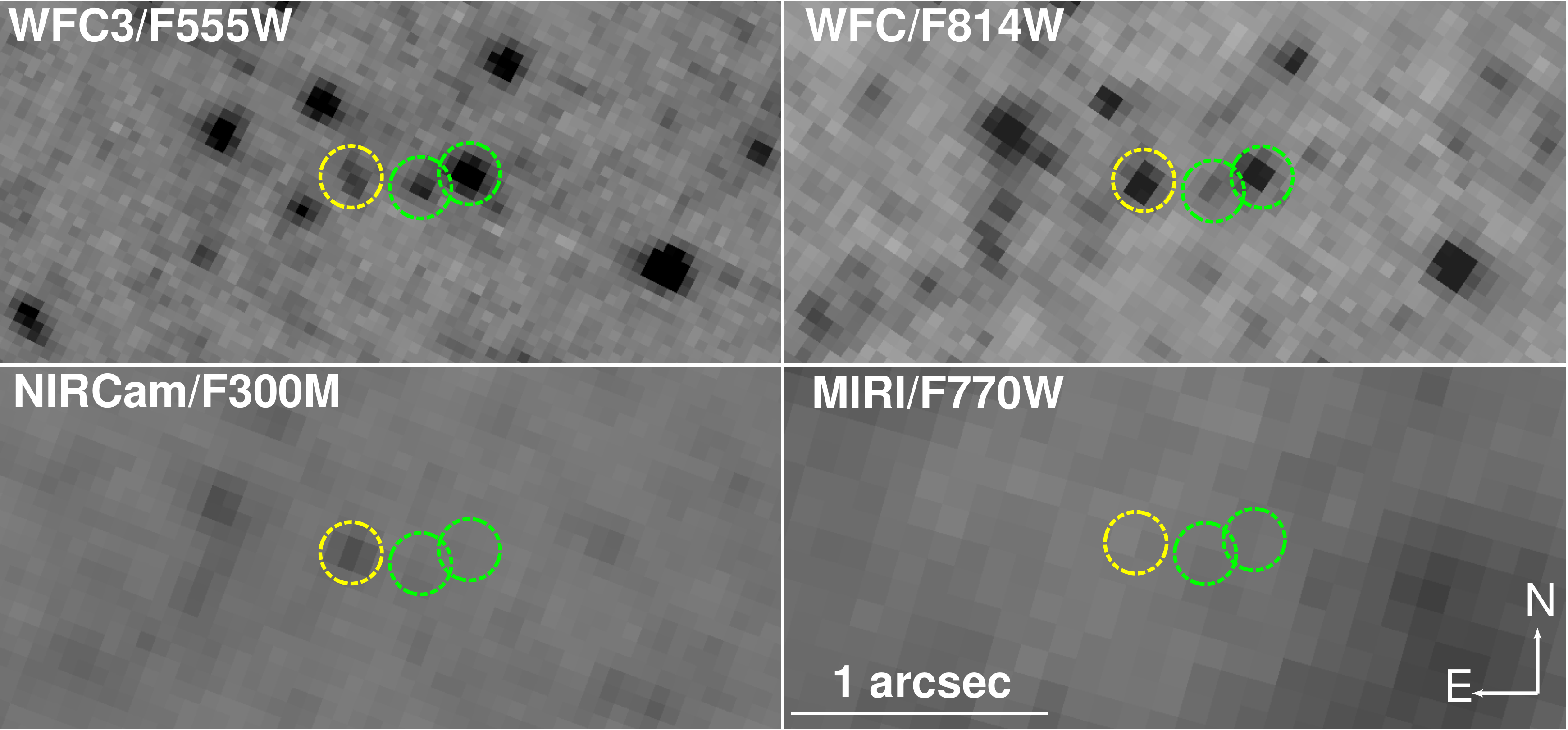}
\caption{The corrected X-ray position of the counterparts of ULX-5 are shown on filters captured bye {\it HST/WFC3} and {\it HST/ACS/FWC}, {\it JWST/NIRCam}, and {\it JWST/MIRI} respectively. Other explanations are the same as in Figure \ref{F:multi}.}
\label{F:ulx5}
\end{center}
\end{figure*}

{\it ULX-5}: A unique possible NIR counterpart was determined for ULX-5. In the case of optical study ({\it 2022b}), two possible optical counterparts ({\it X5$\_1$} and {\it X5$\_2$}) have been detected within error radius 0.21 arcsec (green dashed circles in Figures \ref{F:uniq} and \ref{F:ulx5}). These optical counterparts do not match the source identified in NIR images. As if the astrometric error radius calculated in the study of {\it 2022b} is taken as 0.32 arcsec at 3-$\sigma$, this NIR counterpart can also correspond to the possible optical source. Moreover, the second possible counterpart {\it X5$\_2$} has been detected only in the {\it HST} UV filters. It would be more accurate to say that the ULX-5 has three different donor candidates in total, based on both astrometric error positions. Interestingly, each counterpart has different properties, one is observed only in UV and one in optical, and the other detailed in this study is detected in both optical and NIR wavelengths.\\

{\it ULX-6}: A total of four potential NIR counterparts were identified within the error radius at the 3-$\sigma$ detection threshold. We have reported that the ULX-6 has a unique optical counterpart which can only be detected in the F814W filter, but as seen in Figure \ref{F:multi}, it appears that three sources are lined up in the NIR position of this counterpart. This means that the source is actually a mix of these three sources. \\

{\it ULX-7}: For ULX-7 eight NIR counterparts were identified. In the case of optical study, ULX-7 has two possible optical counterparts. As seen in Figure \ref{F:rgb}, it is located in a highly crowded region in terms of star or star candidate populations, making it very difficult to determine which is the true counterpart.\\

{\it ULX-8}: A unique NIR counterpart was determined within the error radius. Just as with the sources ULX-3 and ULX-4, the optical counterpart(s) were not determined.\\

Considering multiple counterparts, it shows that sources located as far away as $\sim$ 10 pc ($\sim$ 0.12 arcsec) within the 79 pc error radius are distinguishable. Considering the distance of this galaxy NGC 1672, the resolution of the {\it JWST/MIRI} detector is almost 9 pc which is one-quarter of the resolution of the {\it NIRCam} detector. This means that for the {\it JWST/MIRI} detector, to distinguish any counterparts the distance between sources must be at least 40 pc.\\

\subsection{False Positive Rate}

\begin{figure}
\begin{center}
\includegraphics[width=\columnwidth]{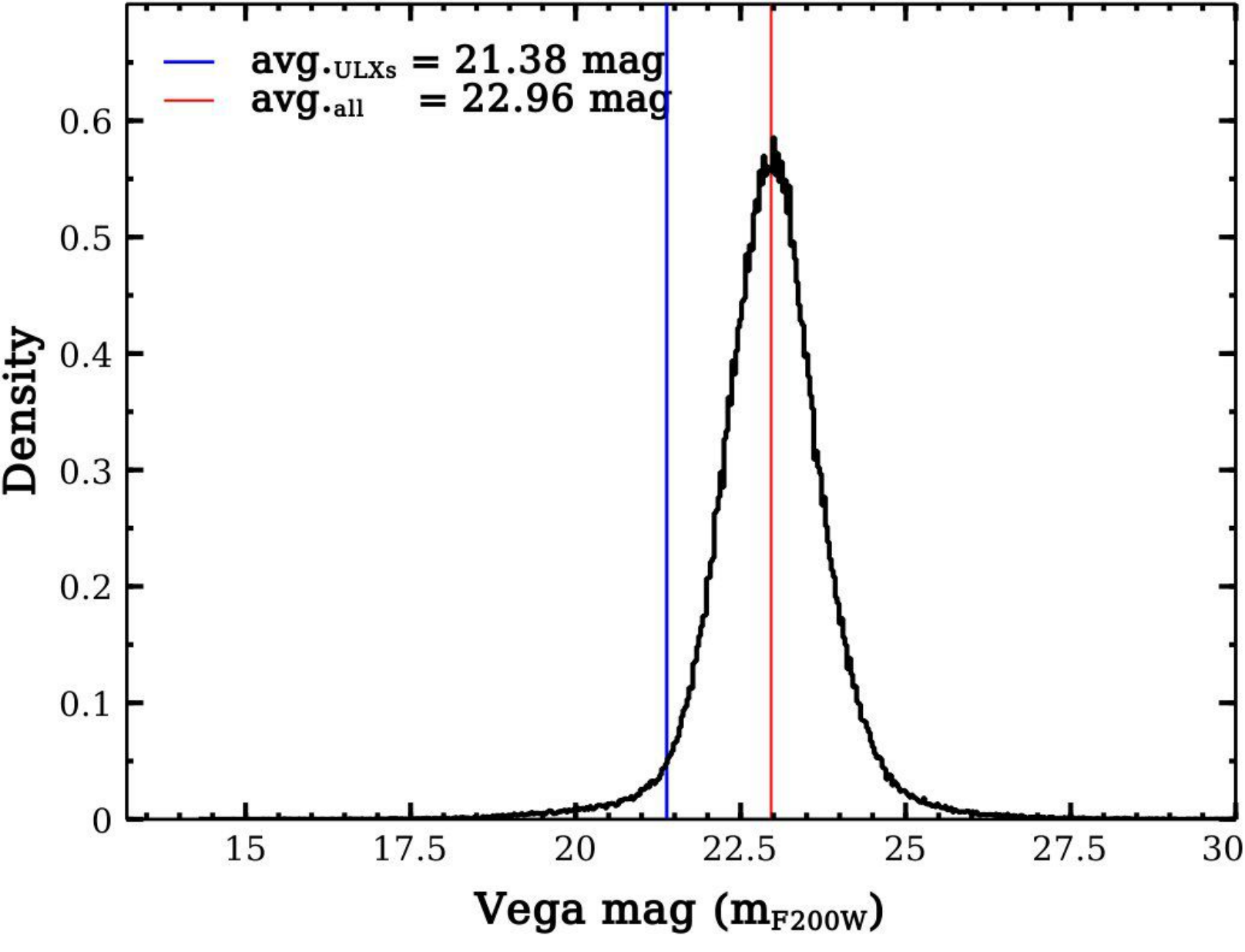}
\caption{The density diagram of NIR sources detected above the 3-$\sigma$ threshold in the {\it JWST/NIRCam} F200W filter. The vertical solid blue and red lines represent the average Vega magnitudes of ULXs and all NIR sources.}
\label{F:is}
\end{center}
\end{figure}

Since the ULXs in NGC 1672 are located in spiral arms (a common feature of ULXs), it is highly likely that we can determine the false positive counterparts. To assess the false positive counterparts (N$_{FP}$), the false positive rates ({\it FPR}) were derived using a similar approach to \cite{2022PASJ...74..283M}. By using Equations \ref{eq2} and \ref{eq3} the number of the \textit{N$_{FP}$} and {\it FPR} were calculated.

\begin{equation}\label{eq2}
N_{FP}= N_{X} \times \pi \times r^2 \times \Sigma
\end{equation}

\begin{equation}\label{eq3}
FPR= 100 \times N_{FP} /  N_{X}
\end{equation}
where N$_{X}$ is the number of ULX, {\it r} $=$ 0.38 arcsec is the
astrometric error radius and as described by \cite{2022PASJ...74..283M} $\Sigma$ is the surface number density of NIR sources, which varies as a function of the magnitude (see the first column in Table \ref{T:fpr}) in {\it NIRCam} filters for NGC 1672. Since this study focuses on unique counterparts, these calculations are considered only for sources ULX-1, ULX-4, ULX-5, and ULX-8. Additionally, the magnitude ranges were chosen according to NIR counterparts which are given in Table \ref{T:fotometri}. As seen in Table \ref{T:fpr}, the average {\it FPR} is $\sim$ 6.5\% and it increases as the magnitude increases in general since the number of NIR sources in galaxy NGC 1672 is larger in these magnitudes ranges (see Figure \ref{F:is}). However, the {\it FPR} values indicate that the NIR counterparts determined in the error radius in this study are reliable. Additionally, as seen in Figure \ref{F:is}, for the galaxy NGC 1672, the lower magnitude limit for the detection of possible NIR counterparts is 27.5 mag at the 3-$\sigma$ detection threshold.

\begin{table*}
\centering
\caption{The false positive rates for detection NIR counterparts}
\begin{tabular}{cccrcccccccccccc}
\hline
Range$^{a}$ & \multicolumn{4}{c}{N$_{X}$$^{b}$}  && \multicolumn{4}{c}{FPR$^{c}$ (\%)} & &  \\
\\
Mag. & F200W & F300M &  F335M & F360M && F200W & F300M &  F335M & F360M\\
\hline
18.62-19.62 & 1     & ...   & 1     & 1     && 2.7 & ... & 2.5 & 2.9\\
19.62-20.62 & ...   & 1     & 1     & ...   && ... & 3.2 & 3.7 & ...\\ 
20.62-21.62 & 1     & ...   & 1     & 1     && 4.2 & ... & 4.4 & 5.1\\
21.62.22.62 & 2     & 2     & ...   & 1     && 7.6 &  6.8 & ... & 9.6\\
22.62-23.62 & ...   & 1     & 1     & 1     && ... & 14.2 & 12.7 & 10.5\\
\hline
\label{T:fpr}
\end{tabular}
\\$^{a}$ Vega magnitude ranges $^{b}$ Number of ULXs $^{c}$ The false positive rate
\end{table*}

\subsection{Spectral Energy Distributions} \label{sed5}

\begin{figure}
\begin{center}
\includegraphics[width=\columnwidth]{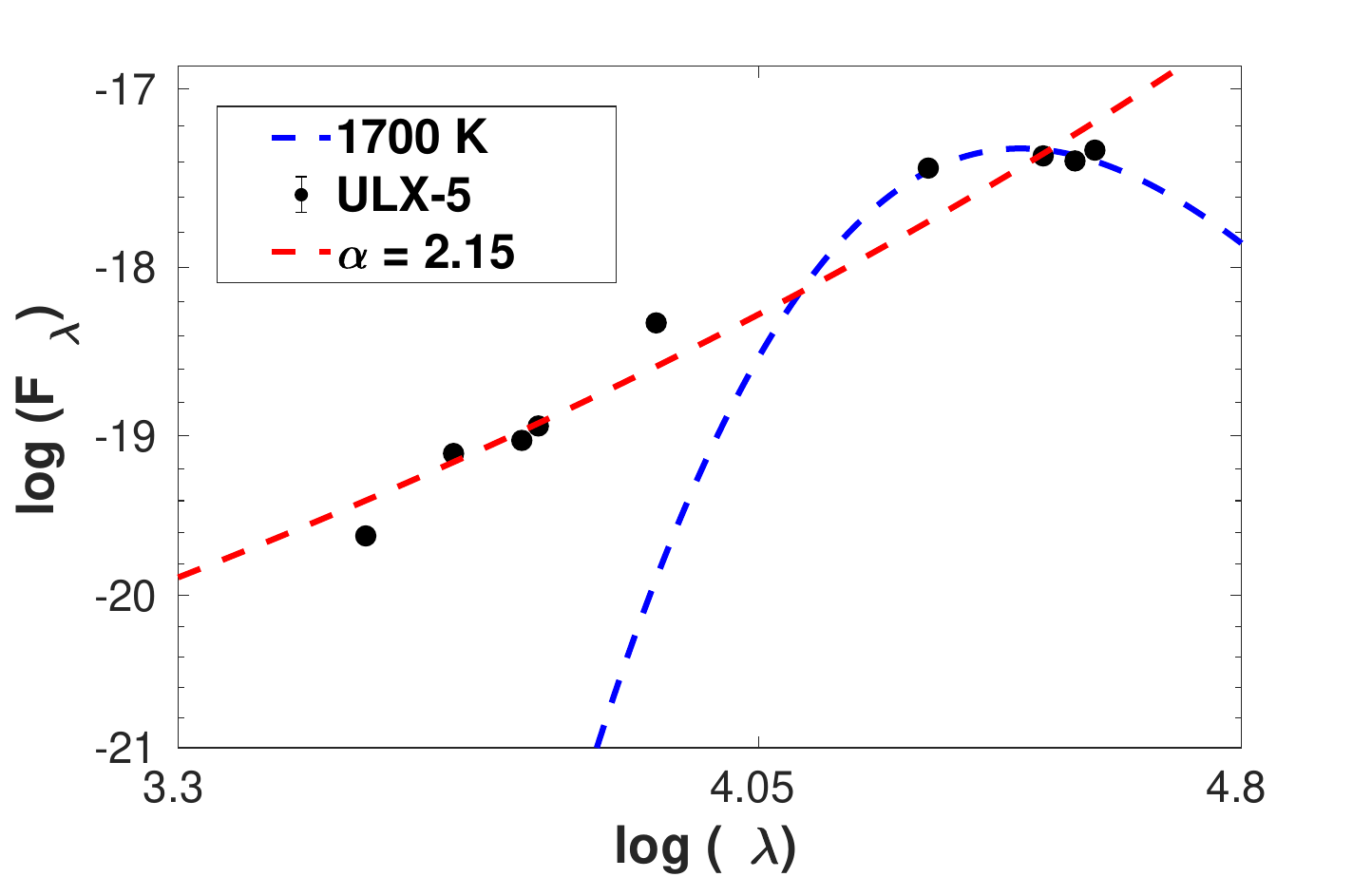}
\caption{The dereddening corrected SED of ULX-5. The blackbody model (1700 K) and power-law ($\alpha$) models are shown by blue and red dashed lines. All data are shown with filled black circles and their respective errors with bars. The units of y and x axes are erg s$^{-1}$ cm$^{-2}$ \AA$^{-1}$ and \AA, respectively.}
\label{F:SED}
\end{center}
\end{figure}

\begin{figure*}
\begin{center}
\includegraphics[angle=0,scale=0.45]{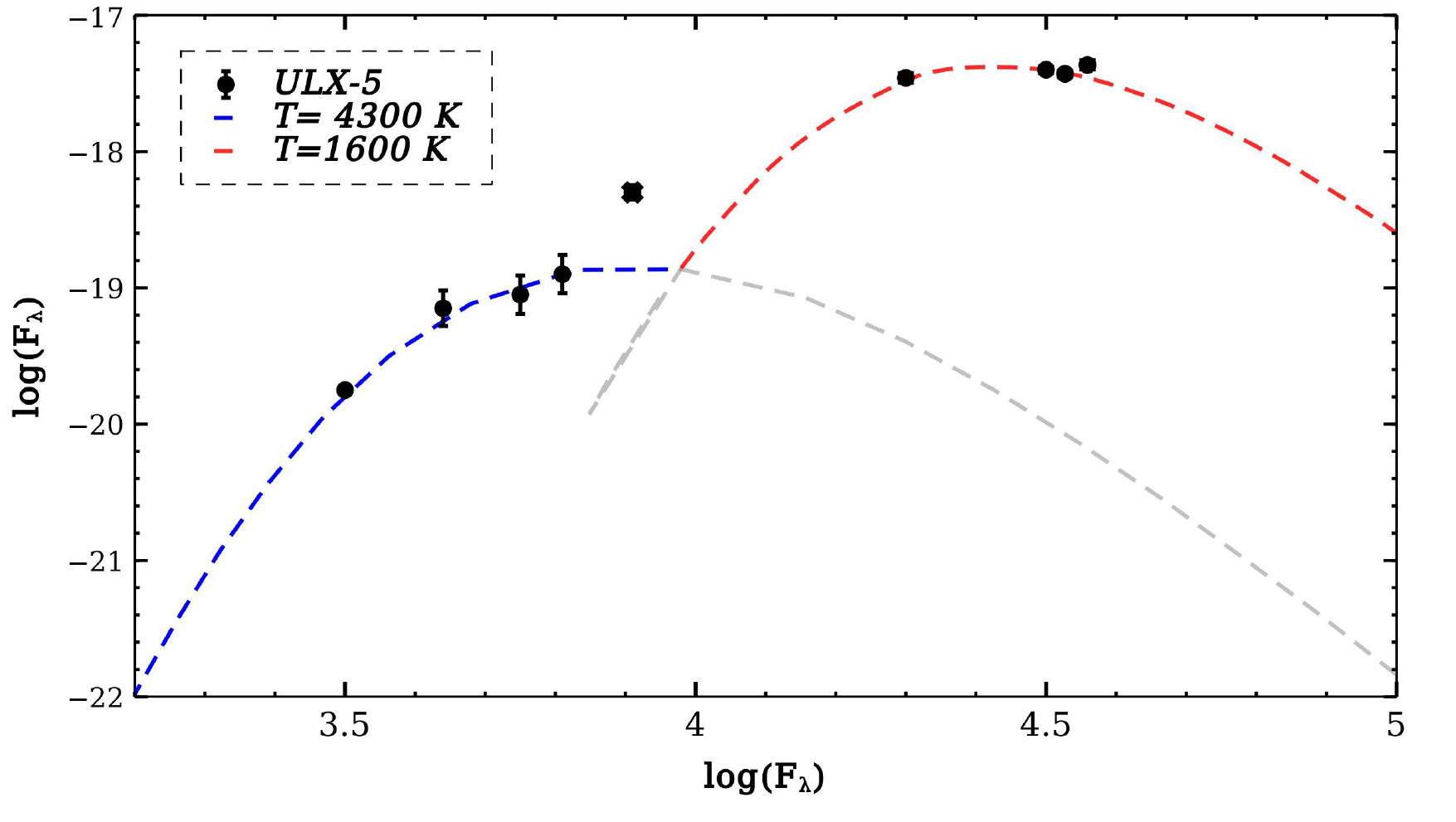}
\caption{The dereddening corrected SED of ULX-5 constructed in Section \ref{sed5}. The dashed blue and red lines show the blackbody models with temperatures of 4300 K and 1600 K. The cross symbol shows F814W observation which is not included in fittings. The units of y and x axes are erg s$^{-1}$ cm$^{-2}$ \AA$^{-1}$ and \AA, respectively.}
\label{F:sed5}
\end{center}
\end{figure*}

\begin{figure}
\begin{center}
\includegraphics[width=\columnwidth]{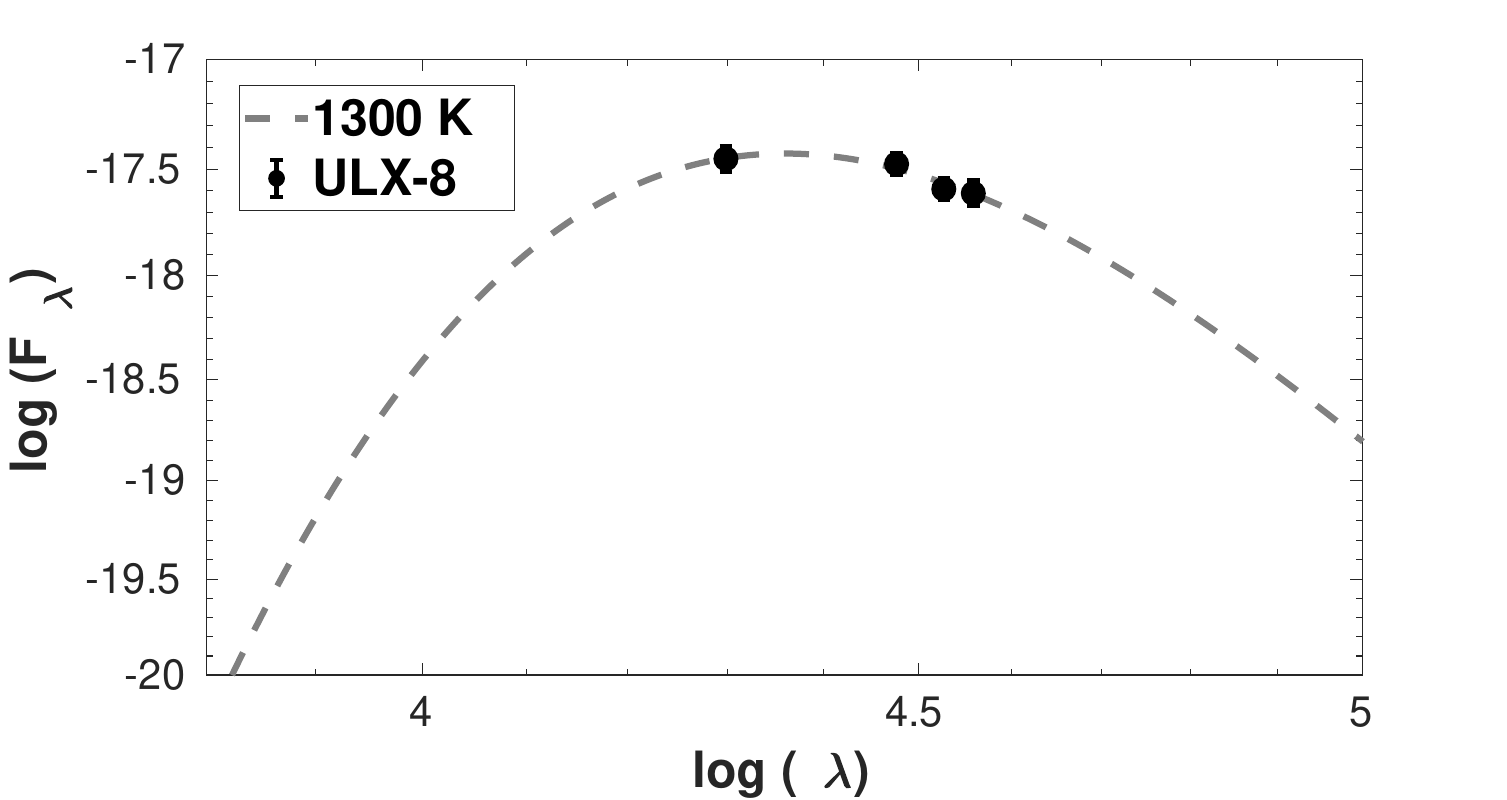}
\caption{The dereddening corrected SED of ULX-8. The gray dashed line shows the blackbody model (1300 K) for ULX-8. All data are shown with filled black circles and their respective errors with bars. The units of y and x axes are erg s$^{-1}$ cm$^{-2}$ \AA$^{-1}$ and \AA, respectively.}
\label{F:SED8}
\end{center}
\end{figure}

The spectral energy distributions (SEDs) of the counterparts were created by utilizing the flux values derived from Table \ref{T:fotometri} to determine the optical and NIR characteristics of the possible donor stars. For this, the SEDs were fitted using either a blackbody or a power-law, F$\propto$ $\lambda^{\alpha}$, spectrum by following the steps in our previous study, {\it 2022b}. The reddening corrected SED of ULX-5 was adequately fitted by the power-law$+$blackbody using \textit{HST} and \textit{NIRCam} data. The best-fitting parameters of the two component model $\alpha$=2.15 $\pm$ 0.11 and blackbody {\it T} = 1700 K $\pm$ 450 at 3-$\sigma$ confidence level. The reduced chi-square, $\chi^2_{\nu}$, value of this best fit is 0.78 with the number of degrees of freedom (dof) of five (see Figure \ref{F:SED}). However, when the {\it HST/WFC3} F814W observation was excluded from the SED, the optical emission of counterpart ULX-5 is well-fitted by the {\it blackbody} {\it T} = 4300 $\pm$ 200 K while NIR emission was well-fitted by {\it blackbody} with 1600 $\pm$ 350 K (see Figure \ref{F:sed5}. The $\chi^2_{\nu}$ for this two-component model is 0.91 with the dof of four at 3-$\sigma$ significance. In the case of ULX-8, the SED was well-fitted by the blackbody spectrum with $\sim$ 1300 K. For this fitting, $\chi^2_{\nu}$ is 0.87 with dof=2 at 3-$\sigma$ level. The dereddening corrected SED ULX-8 is shown in Figure \ref{F:SED8}. For the remaining sources, ULX-1 and ULX-4, an acceptable model could not be obtained.\\

\section{Results and Discussion} \label{sec:4}

\subsection{Astrometry and Identification of Counterparts}

There is no dramatic difference found between the astrometric calculations in our previous optical study and in this study. This indicates that the true counterparts of ULXs at different wavelengths were investigated. Taking into account both calculations, for each ULX, ULX-2, ULX-3, ULX-6, and ULX-7, at least four NIR counterparts were identified within the astrometric error radius of 0.38 arcsec, and except for ULX-3, at least one of these counterparts matches the optical counterpart. Since there are too many possibilities to decide which is a possible donor star, more data and simultaneous X-ray and multi-wavelength observations are needed to identify possible donor stars. Therefore, only ULX-1, ULX-4, ULX-5, and ULX-8 which have unique NIR counterparts are detailed in this study. For the NIR counterpart of ULX-5, an optical counterpart was identified which is particularly faint at optical wavelengths, while for the NIR counterpart of ULX-1, an optical counterpart was identified which can only be detected in the F814W filter. For ULX-4 and ULX-8, no optical counterpart was identified within the astrometric error radius. \\

\begin{figure*}
\begin{center}
\includegraphics[angle=0,scale=0.3]{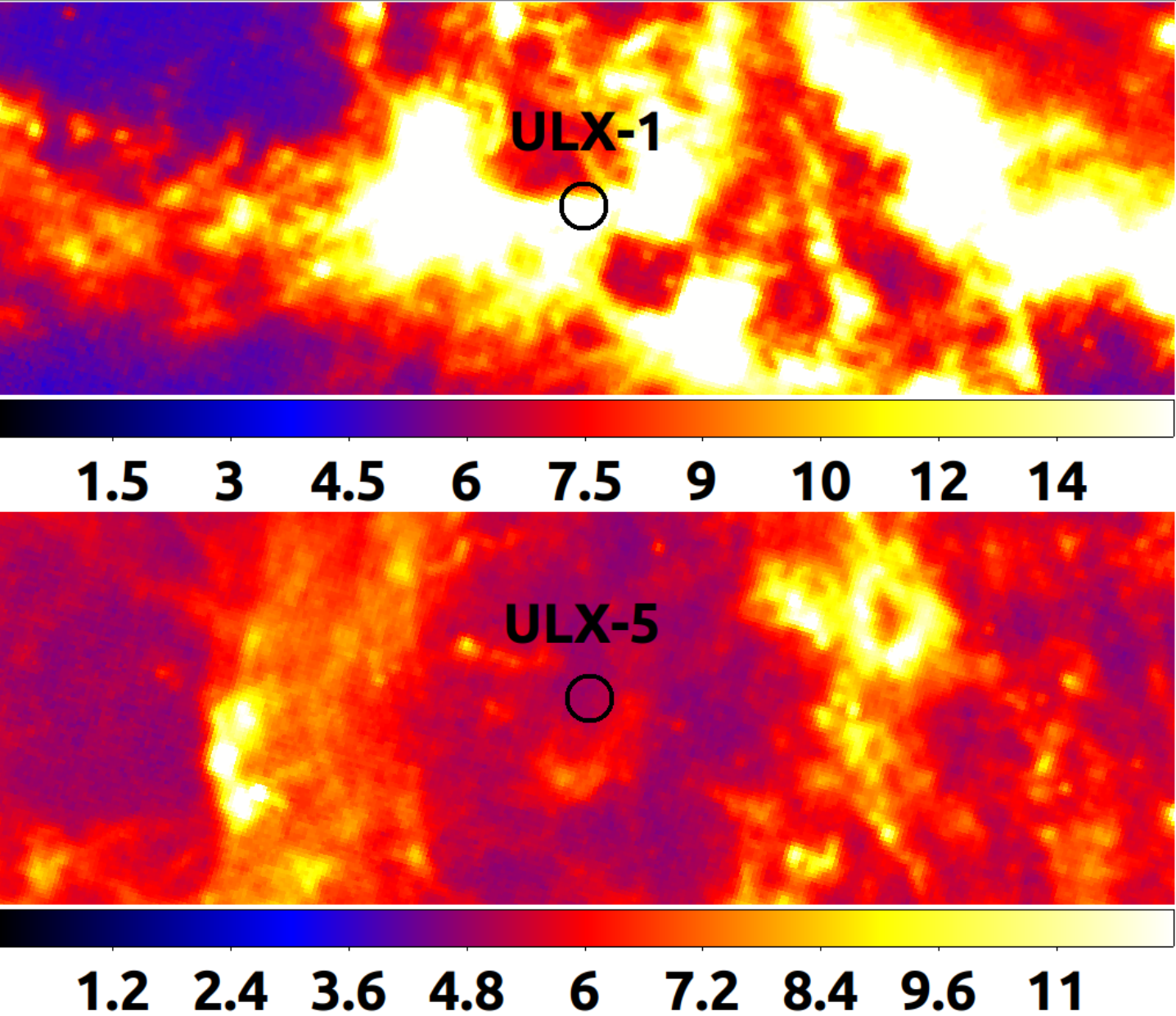}
\includegraphics[angle=0,scale=0.3]{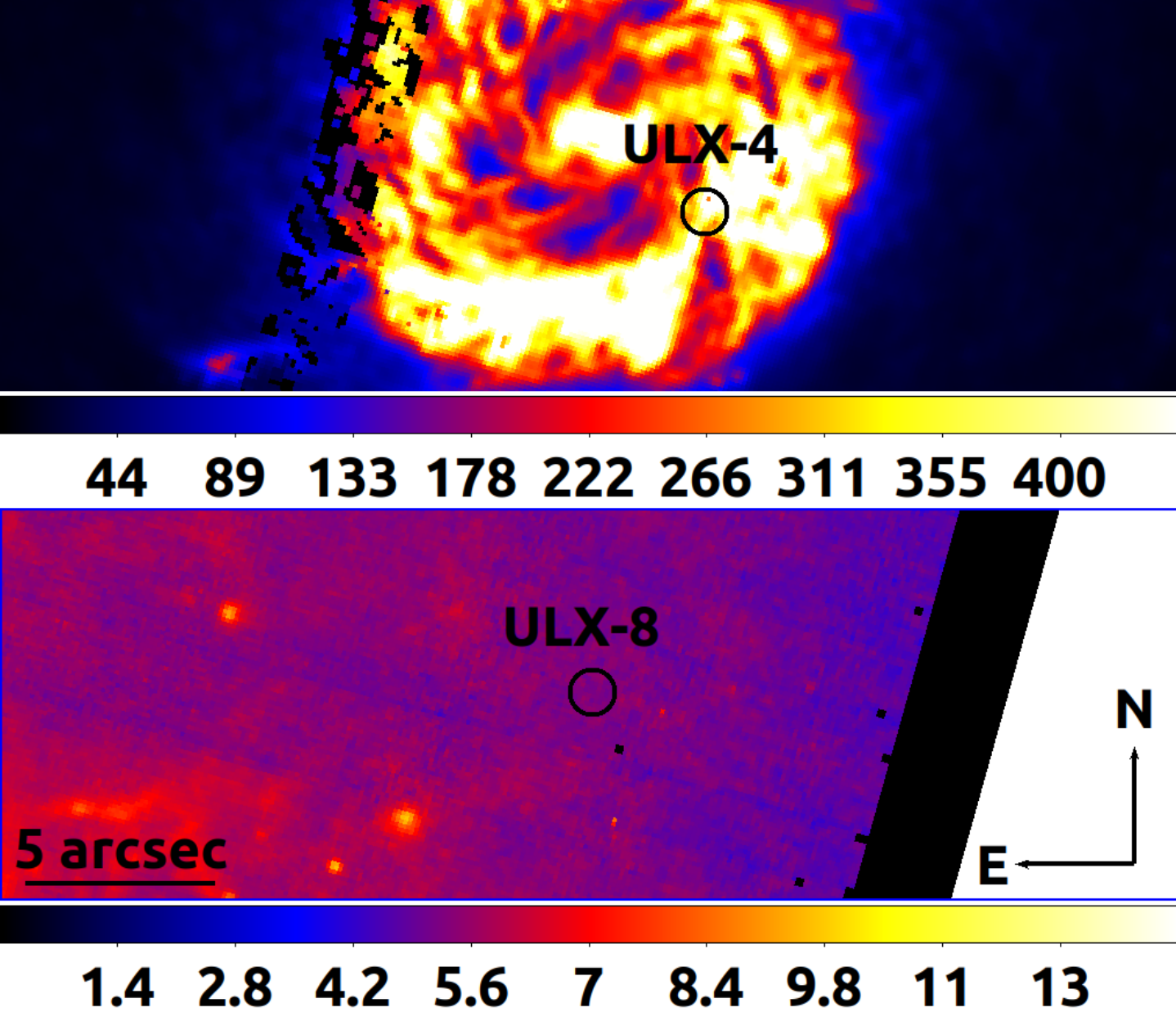}
\caption{The locations of the ULX-1, ULX-4, ULX5, and ULX-8 (solid unfilled black circles) on the colored {\it JWST/MIRI} F770W images. The numbers shown for the color bars are in units of MJy/sr. All panels are the same scale and the north is up, east is left.}
\label{F:or}
\end{center}
\end{figure*}

As seen in Figure \ref{F:or}, locations of ULX-1, ULX-4, ULX5, and ULX-8 are displayed on the {\it JWST/MIRI} F770W image, especially ULX-4 is located in very dense fields. These dense fields might scatter optical radiation, thus it is highly likely that only the NIR counterpart was detected. We may have observed the radiation in the NIR bands more dominantly due to the circumbinary disk/dust surrounding ULXs. Furthermore, no counterpart(s) for these four ULXs could be detected in the {\it JWST/MIRI} detector (7.7 - 21 $\mu$m). This may be due to the fact that the {\it JWST/MIRI} detector does not have sufficient spatial resolution to resolve sources at distances of 16.3 Mpc. \\

\begin{figure}
\begin{center}
\includegraphics[width=\columnwidth]{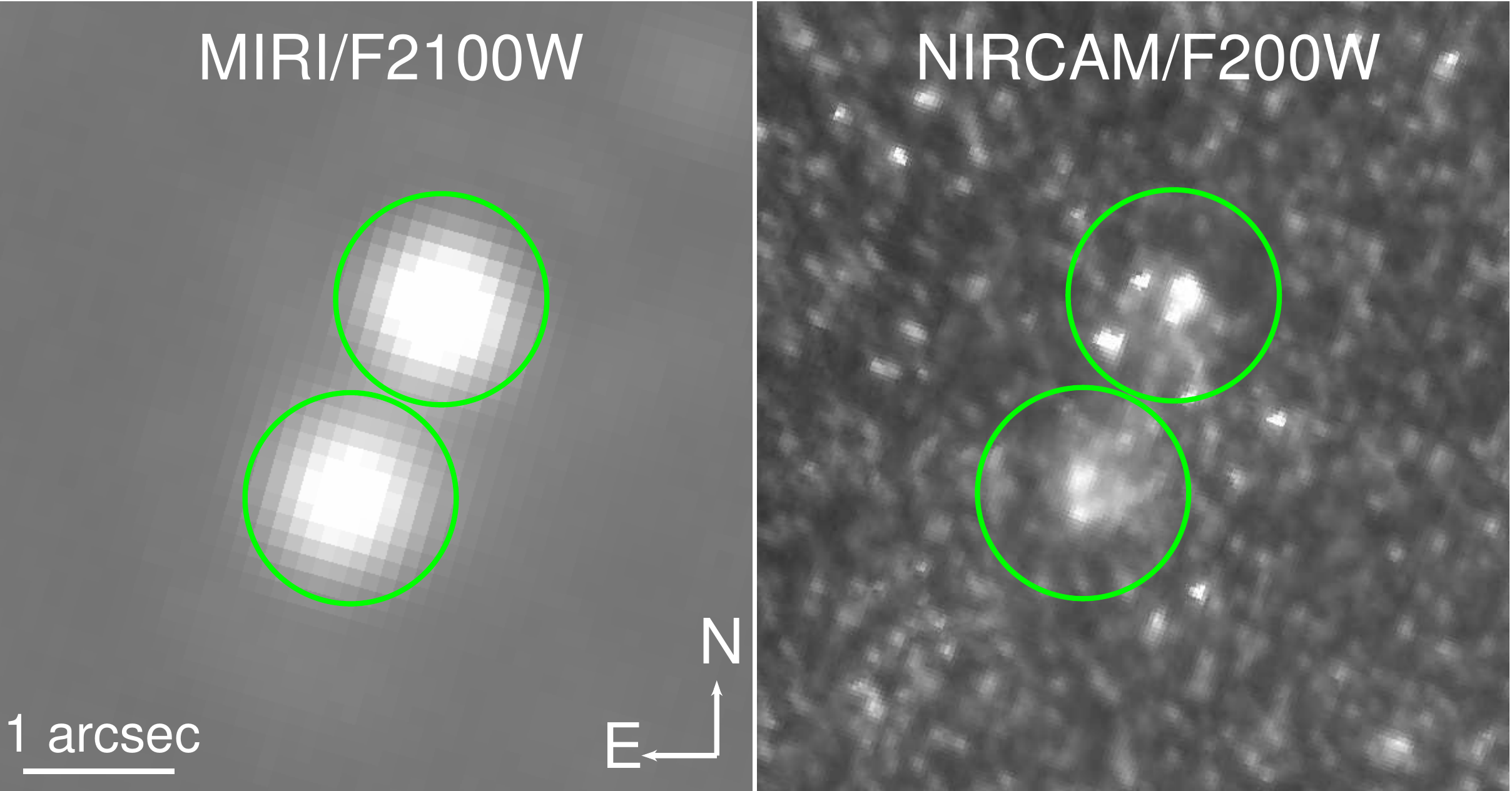}
\includegraphics[width=\columnwidth]{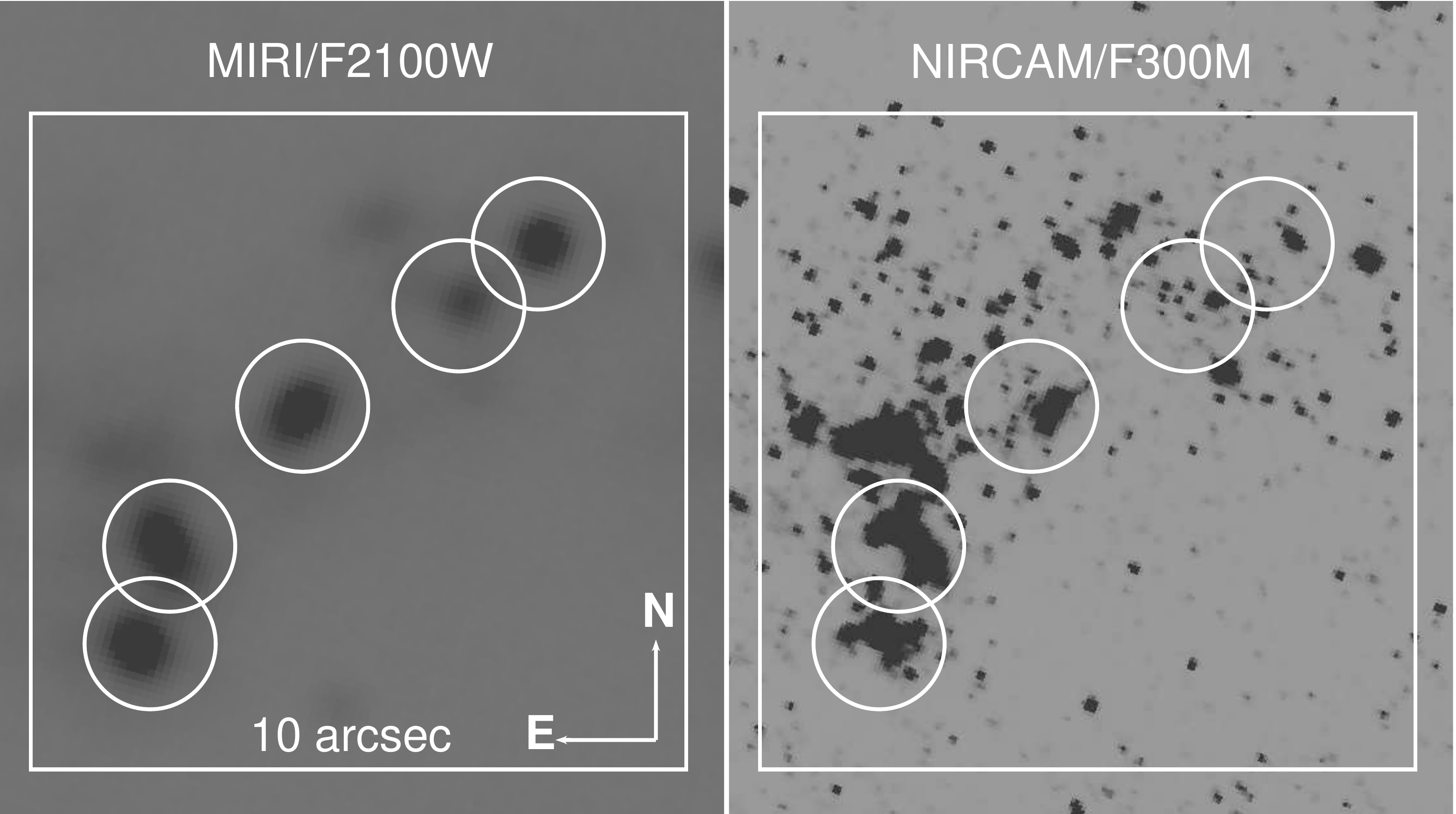}
\caption{Upper two panels: Comparison of {\it JWST/MIRI} F2100W image (0.11 arcsec/pixel) (left) with {\it NIRcam} F200W image (0.03 arcsec/pixel) (right). Comparisons of resolution for F2100W is 0.111 arcsec/pixel and for F200W is 0.031 arcsec/pixel. Lower two panels: Comparison of {\it JWST/MIRI} F2100W image (left) with {\it JWST/NIRCam} F300M image (0.063 arcsec/pixel) (right).}
\label{F:compare}
\end{center}
\end{figure}

Since ULXs are often found in crowded fields, such as spiral arms, it is quite possible that in such distant galaxies, we may identify blended sources as counterparts. As seen in Figure \ref{F:compare}, randomly selected sources that can be resolved in the {\it JWST/NIRcam} images (0.03 arcsec/pixel) are observed as bright sources in the {\it JWST/MIRI} images (0.11 arcsec/pixel). If the ULXs in galaxy NGC 1672 were observed with ground-based and/or {\it Spitzer/IRAC}\footnote{https://irsa.ipac.caltech.edu/data/SPITZER/docs/irac/} infrared detectors, it is highly likely that the multiple counterparts shown in Figure \ref{F:multi} would be identified as the bright unique sources. More detailed comparisons such as {\it JSWT} and {\it Spitzer/IRAC} were given in the study of \cite{2023ApJ...944L..17L}. In conclusion, the advanced capabilities of {\it JWST} will facilitate a better understanding and accurate characterization of distant ULXs. In light of this, it is necessary to revisit the results obtained from previous IR observations and conduct new observations with {\it JWST} to gain further insights into the nature of ULXs especially in distant galaxies.\\

\subsection{Photometric Results of Counterparts}

The notable multi-wavelength characteristics of each ULX which have unique NIR counterparts were presented as follows:\\

\subsubsection{ULX-1 and ULX-4}

Reddened Vega magnitudes of this NIR counterpart of ULX-1 in each filter are given in Table \ref{T:fotometri}. The wavelength range of JHK NIR bands is 1.02-2.39 $\mu$m so there is no other choice but to compare the {\it NIRCam} F200W ($\sim$ 2 $\mu${\it m}) with the K-band. Based on the distance NGC 1672 of 16.3 Mpc the absolute magnitude M$_{F200W}$ of NIR counterparts of ULX-1 are $\sim$ -9.4 mag. Moreover, the F814W-F360W filters were taken into account as colors to qualify the sources as red or blue. The intrinsic color of this counterpart is as F814W-F360W $\sim$ $-$ 3.2 mag thus, this source may have red color enough to be an RSG or RG (red giant). For the RG scenario, the probability of detecting a low-mass donor star at this distance is very low. On the other hand, the low-mass late-type RGs, such as Ultra-compact X-ray binaries such as reported by \cite{1984ApJ...283..232R} and \cite{1986ApJ...311..226N} may be powered by the accretion disk therefore, it might have behaved as a high-mass X-ray binary (HMXB) (due to the contribution of X-rays). The presence of H$\alpha$ emission can be reddened sources \citep{2016ApJ...824...71C,2023MNRAS.519.4826A} thus, may make it appear the fainter magnitudes of counterparts. For NGC 1672, the cataloged H$_{II}$ regions taken from \cite{2022A&A...662L...6B} were compared to the counterparts, but no sources matching the H$_{II}$ regions were found. However, we reported in the study of {\it 2022b} that only ULX-1 could be in an uncatalogued H$_{II}$ region. As seen in Figure \ref{F:or}, ULX-1 is located in a very dense field which might scatter optical radiation. Therefore, the optical counterpart ({\it X1$\_2$}) of NIR source may not be detected. Moreover, we may have observed the radiation in the NIR bands more dominantly due to the circumbinary disk/dust surrounding ULX-1.\\

\begin{figure*}
\begin{center}
\includegraphics[angle=0,scale=0.5]{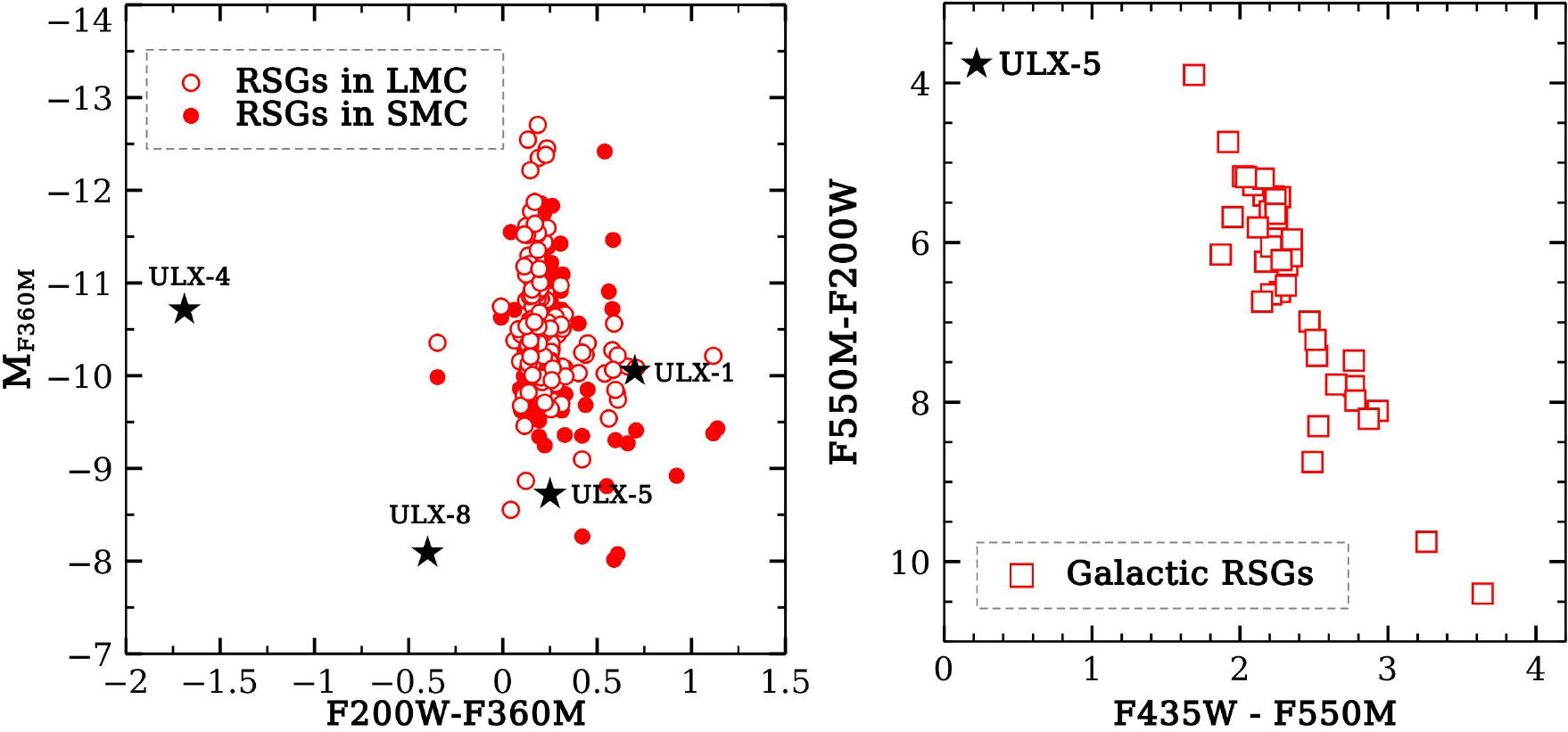}
\caption{Left panel: The color-magnitude diagram of counterparts ULX-1, ULX-4, ULX-5, and ULX-8 (filled black stars) plotted over RSGs in the LMC (unfilled red circles) and SMC (filled red circles) with well-known spectral type compiled by \citep{2011ApJ...727...53Y} and \citep{2012ApJ...754...35Y}. Left panel: two-color diagram of counterpart of ULX-5 plotted over Galactic RSGs (unfilled red squares) \citep{2005ApJ...628..973L}.}
\label{F:RSG}
\end{center}
\end{figure*}

The absolute $\sim$ K-band magnitude of the NIR counterpart is consistent for some of the M-type RSGs ($-$ 8 to $-$ 11 mag, \citealp{1985ApJS...57...91E}), but is not sufficient to identify them as RSGs. The color (K-[3.6]) and absolute [3.6]-band ($\sim$ F360M) magnitude values of RSGs in Large Magellanic Cloud (LMC) \citep{2011ApJ...727...53Y} and Small Magellanic Cloud (SMC) \citep{2012ApJ...754...35Y} were plotted and compared with the ULXs in this study. As seen in the left panel of Figure \ref{F:RSG}, ULX-1 is on the borderline of the colors of RSGs, which does not make it a strong RSG candidate, but this study does not exclude the possibility that it could be an RSG. Finally, for the optical-UV-dominant counterpart of ULX-1, we previously reported hints that its optical emission comes mostly from the accretion disk. As a result of both optical and NIR analyses, it has two possible donors, either optical-UV dominant or NIR dominant. \\

As seen in Table \ref{T:fotometri}, ULX-4 has apparent magnitudes highly dependent on wavelengths. The intrinsic color of ULX-4 is as F200W-F360M $=$ $-$1.7 mag and absolute magnitudes M$_{F200W}$ is $\sim$ $-$12.4 mag. While RSGs are the most luminous sources in star-forming galaxies, RGs (red giants) are the highly luminous sources in passive galaxies, and also in their later evolution, they are the dominant sources in the NIR bands \citep{2007A&A...468..205L}. However, as seen in Figure \ref{F:RSG}, it does not have the colors and luminosities to be an RSG. In this work and also our previous optical study, no optical counterparts were found for ULX-4 within the respective error radii. The scenarios suggested for ULX-1 may also be consistent for ULX-4, but, because of its high magnitude (M$_{F200W}$ $\sim$ $-$12 mag), the possibility of being an AGN or star cluster should not be excluded. Although to constrain the age of the counterparts, the PARSEC \citep{2012MNRAS.427..127B} isochrones are a good tool, a model consistent with the colors of ULX-1 and ULX-4 could not be obtained. In addition, CMDs are pretty work for stellar populations but, since ULX-4 is more likely to be an AGN or star cluster, stellar isochrones may not have been determined for this source. In order to constrain the nature of counterparts, SEDs also were plotted but no acceptable model (blackbody and/or a power-law) were found. More observations are needed for the nature of both ULX-1 and ULX-4.\\

\subsubsection{ULX-5}

The SED (optical to NIR) for the counterpart of ULX-5 is adequately fitted with two component power-law $+$ blackbody model with $\alpha$ $=$ 2.15 $\pm$ 0.16 and T $\sim$ 1700 K. As seen the reddening corrected SED of ULX-5, is displayed in Figure \ref{F:SED}, optical data is represented by the power-law model, while NIR data is represented by the blackbody model. The power-law model can be observed in case optical emission originates mostly from accretion disks \citep{2011ApJ...737...81T}, in general, blackbody emissions are expected to come from donor stars. However, the spectrum shows an excess over the power-law spectrum above optical wavelengths in the F814W band. Furthermore, the power-law does not cut-off at optical wavelengths. This indicates that the emission of ULX-5 could be due to the accretion disk, donor star, or circumbinary disk/dust. In the case of the accretion disk, such an emission indicates the presence of a larger ($\geq$ 10$^{13}$ cm) disk than expected for ULXs \citep{2011ApJ...737...81T,2012ApJ...745..123G,2014MNRAS.444.2415S}. Therefore, this SED is not appropriate for determining the nature of ULX-5.\\

When the F814W observation was not included, the spectrum of ULX-5 was well-fitted by two blackbody models with significantly better fitting statistics. In this case, temperatures of 4300 K and 1600 K were found for optical and NIR wavelengths, respectively. As seen in Figure \ref{F:sed5}, the blackbody model representing optical radiation is cut-off at the end of the optical wavelength and clearly an NIR excess is observed. Since this SED is more reasonable both in terms of fitting statistics and physical parameters, the nature of ULX-5 is interpreted according to this SED. According to the results obtained from this SED, the observed radiation could be either the outer part of the accretion disk (irradiated), the possible donor star, the circumbinary disk, or the variable jet \citep{2016ApJ...831...88D,2019ApJ...878...71L}. The F814W observation is not simultaneous with the NIR observations, and the variability feature of this counterpart is unclear. Hence, the position of the F814W observation in the SED might be due to the variability at long wavelengths. In this case, the variable jet scenario may also make sense for ULX-5, but there is neither evidence nor sufficient data to explain it. The other three scenarios are detailed in the following:\\

{\it The Outer Part of the Accretion Disk:} In typical Galactic black hole binaries, the X-ray flux from the inner source irradiates and gets reprocessed within the outer regions of the accretion disk, often exerting a dominant effect on the spectrum observed in UV and optical wavelengths. Additionally, counterparts of ULXs are dominated by emission from the accretion disk with intrinsic and/or reprocessed. By following study \cite{2011ApJ...737...81T}, the surface area of the emitting regions of ULX-5 were derived for using blackbody temperatures of 4300 K and 1600 K as 8 $\times$ 10$^{24}$ {\it cm$^{2}$} and 8 $\times$ 10$^{26}$ cm$^{2}$ or a radii of 3 $\times$ 10$^{12}$ and 3 $\times$ 10$^{13}$ {\it cm}, respectively. These surface areas of the emitting regions, especially for the NIR emission are larger than expected for ULXs \citep{2012ApJ...745..123G,2014MNRAS.444.2415S}. On the other hand, based on irradiated disk models of ULXs, larger outer disk radii of $\sim$ 10$^{12}$ {\it cm} have been reported by a few authors (e.g., \citealp{2011ApJ...737...81T,2012ApJ...745..123G,2012ApJ...750..110T}). However, the radius calculated for 1600 K is too large to be physically acceptable hence, this study rules out the scenario in which NIR emission is powered by an outer accretion disk and donor star. In this case, for NIR excess, the circumbinary disk or warm dust scenarios are more acceptable, while for optical radiation, a circumbinary disk or dust-affected donor star scenario is more reasonable.\\

{\it The Possible Donor Star}: The M$_{F200W}$ and M$_{F555W}$ magnitudes NIR counterpart of ULX-5 are $\sim$ $-$ 8.5 mag and $\sim$ $-$ 4.5, respectively. The color between F550M and F360M of counterpart was derived as $\sim$ 4.2 mag this indicates that the source is highly red. The optical counterparts of ULX-5 can be considered RSG based on effective temperatures of late-type (K-M) Galactic SRGs ($\sim$ 3660 K, \citealp{2005ApJ...628..973L}). Moreover, as seen in the color-magnitude diagram (see Figure \ref{F:RSG}) it is located borderline population of some of the RSGs in SMC \citep{2012ApJ...754...35Y} and LMC \citep{2011ApJ...727...53Y}. Additionally, to determine the type of possible donor star of ULX-5, it was compared with 40 out of 74 Galactic RSGs \citep{2005ApJ...628..973L} which have B-V and V-K colors (in the case of this study F550M-F200W vs F435W-F550M). As seen in the right panel of Figure \ref{F:RSG}, since this counterpart is fainter than these Galactic RSGs, especially at short wavelengths, it is placed differently from the RSGs in the two-color diagram. We know that the V-band Vega magnitudes (m$_{V}$) of the optical counterparts are faint, m$_{V}$ $\geq$ 18 mag, and the environments they are in especially interstellar and/or local gas and dust make it difficult for us to observe them \citep{2019ApJ...878...71L}. Since the wavelength range of optical emission is generally the same size as local dust particles, it is easily scattered by the dust while longer-wavelength emission passes through unblocked. Therefore, the optical counterpart of ULX-5 might be observed as very faint ($\sim$ 28 mag) in the {\it HST/WFC3} F336W as well as {\it HST} V-band image that is, it has almost the predicted limiting magnitude of {\it HST/WFC3}. As a result, there is a high probability that the counterpart of ULX-5 is an RSG with K-M spectral type affected by the possible presence of a circumbinary disk or dust around the ULX-5 system. The effects of a circumbinary disk on XRBs with RSG donor star systems are quite possible \citep{2019ApJ...876L..11C}.\\

{\it The Presence of Circumstellar/circumbinary Disk/Dust}: The temperature of 1600 K determined in the NIR SED of ULX-5 is neither hot enough to come from a donor star nor the result of reprocessing from an accretion disc. Since the emissions of the IR wavelengths are less affected by dust extinction the main reason why the optical counterpart of ULX-5 is highly faint except the red filter (F814W) could be due to circumbinary disk/dust. By using a similar approach \cite{2019ApJ...878...71L}, the equilibrium temperature radius provides an estimate of the lower limit or inner radius of
the surrounding warm dust, R${eq}$ derived by Equation \ref{E6}:
\begin{equation}\label{E6}
R_{eq} = \frac{Q_{abs}L_{IR}}{Q_{e}16\pi \sigma T^{4}}
\end{equation}
where {\it Q$_{abs}$} and {\it Q$_{e}$} are the grain absorption and emission Planck mean cross sections, {\it $\sigma$} is Stefan-Boltzmann constant, L$_{IR}$ is the luminosity for RSG in SN2010da/NGC 300 ULX-1 \citep{2019ApJ...883L..34H} with 3900 K. In the case of radiative dust, heating is dominated by the stellar component, {\it L$_{\star}$} = {\it L$_{IR}$}. The value of {\it Q$_{abs}$}/{\it Q$_{e}$} are $\sim$ 2.5. {\it T} is a temperature of 1600 K obtained from the NIR SED of ULX-5. The R${eq}$ was derived as $\sim$ 0.3 AU (astronomical unit) and in the case of the X-ray contribution (5 $\times$ 10$^{5}$ L$\odot$), R${eq}$ was found as $\sim$ 35 AU. We can understand further insight into the circumstellar dust geometry such as estimating the binary separation, {\it a}. The separation value was derived as {\it a}= $\sim$ 3.4 AU\footnote{In this case, the masses of the compact object and donor stars were assumed as 2.5 M$\odot$ and 10 M$\odot$, respectively, and also {\it R$_{L}$} = {\it R$_{\star}$} was accepted. Moreover, the radius {\it R$_{\star}$} was taken as $\sim$ 310 R$\odot$ for RSG in SN2010da/NGC 300 ULX-1 \citep{2019ApJ...883L..34H}.} from the Roche lobe radius {\it R$_{L}$} \citep{1983ApJ...268..368E} Equation \ref{E7}.
\begin{equation}\label{E7}
\frac{R_{L} }{a} = \frac{0.49q^{2/3}}{0.6q^{2/3}+log(1+q^{1/3})}
\end{equation}

If we ignore the X-ray contribution, {\it a} > R$_{eq}$ (0.3 AU), the dust is not likely in a circumbinary disk thus, it is likely due to warm dust or circumbinary dust around the ULX-5. On the contrary, if we take into account the contribution of X-rays, {\it a} < R$_{eq}$ (35 AU), the dust is likely in a circumbinary disk/torus \citep{2019ApJ...878...71L}. Considering all possible scenarios, it is evidence of warm dust destroyed by X-ray emission around the highly probable late-type RSG donor star. High-resolution (0.1 arcsec/pixel) spectral observations such as {\it JWST} Near Infrared Spectrograph ({\it NIRSpec}) (0.6–5.3 $\mu$m) are needed to constrain the nature of possible donor star of ULX-5. The ULX-5 is a transient X-ray source that also exhibits a high degree variability factor of 30 in X-ray of the long-term light curve ({\it 2022b}) but it is not clear whether it is variable at multi-wavelengths. Therefore, simultaneous multi-wavelength and X-ray observations will place stronger constraints on the observed optical and NIR emission for the ULX-5 system.\\

\subsubsection{ULX-8}
The NIR counterpart of ULX-8 has similar magnitudes in almost all NIR filters and its absolute magnitude M$_{F200W}$ is $\sim$ $-$ 8.5. Moreover, color F200W-F360M was derived as $\sim$ -0.4 mag which indicates that it is the relatively fainter source at longer wavelengths in NIR observations but,  the optical counterpart of ULX-8 was not determined. Additionally, it is a persistent X-ray source with a variability factor of $\sim$ 4 (see {\it 2022b}). The NIR SED of ULX-8 is well-fitted by the blackbody model with T $\sim$ 1300 K (see Figure \ref{F:SED8}). The power-law model can be observed in case optical emission originates mostly from accretion disks \citep{2011ApJ...737...81T}, in general, thermal emissions are expected that come from donor stars. However, the temperature obtained from the NIR SED of ULX-8 is not sufficiently hot enough to come from a possible donor star. Therefore, the scenarios recommended for the ULX-5 NIR counterpart are also compatible with this source (e.g., {\it a} > R$_{eq}$) On the other hand, the type of the possible donor star is highly uncertain in the case of these observations, as the NIR radiation is due to dust disturbed by X-rays. On the other hand, it is also possible that the optical counterpart(s) of ULX-8 could be low mass X-ray binary (LMXB) and could not be detected in the astrometric error radius due to being intrinsically faint at the distance of galaxy 1672. Additionally, since ULX-8 is located in very dense dust, therefore, even if it has optical counterpart(s) they may not be detected.\\

\begin{figure}
\begin{center}
\includegraphics[width=\columnwidth]{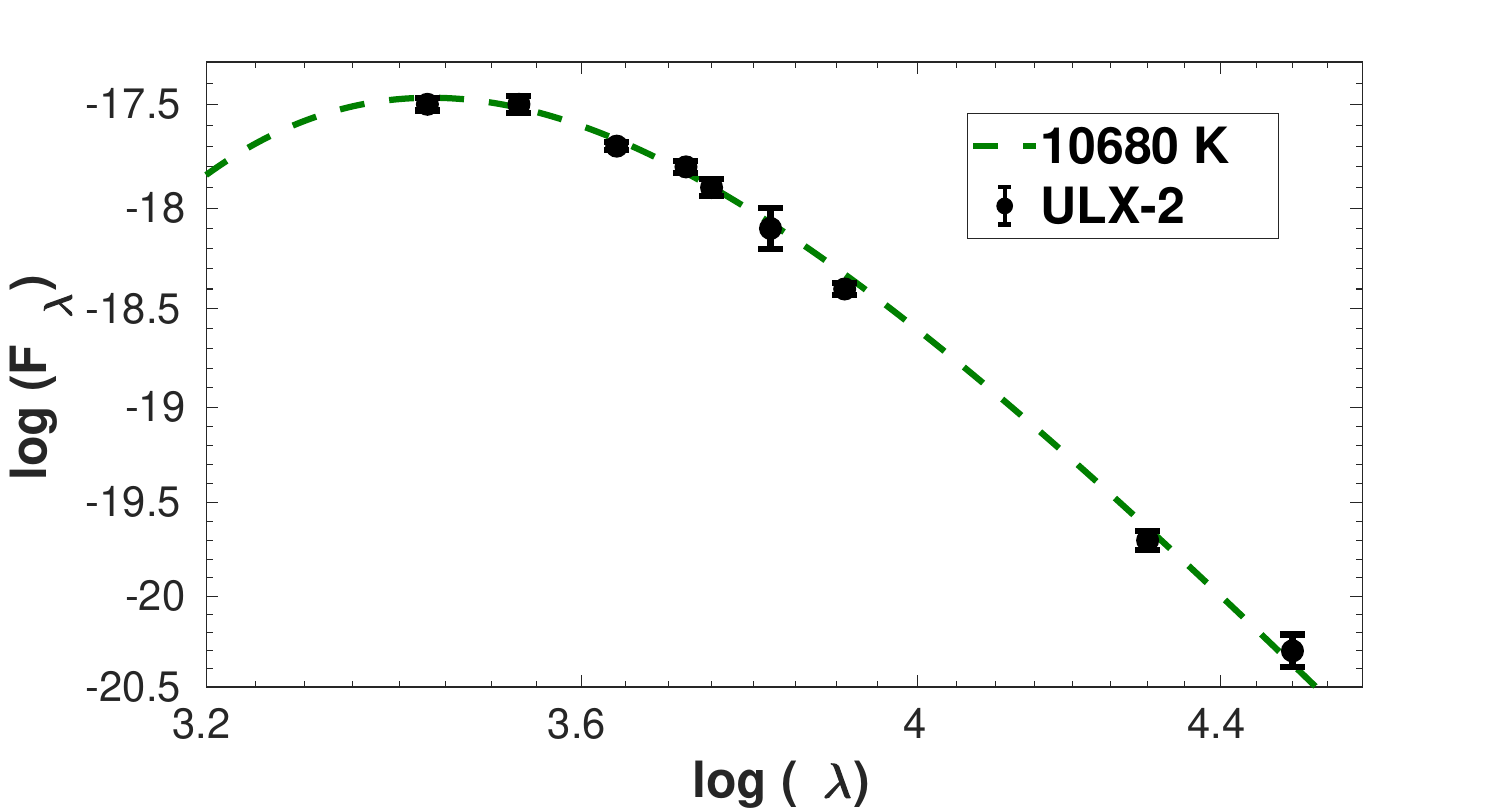}
\caption{The dereddening corrected SED of ULX-2. The green dashed line shows the blackbody model (10680 K) model. Other explanations are the same as in Figure \ref{F:SED8}.}
\label{F:SED2}
\end{center}
\end{figure}

Finally, a special parenthesis for ULX-2, in our previous study, we reported that the SED of its unique counterpart was well represented by a blackbody model with 10000 K. The SED of ULX-2 was re-plotted using fluxes 2 $\times$ 10$^{-20}$ and 5 $\times$ 10$^{-21}$ erg cm$^{-2}$ s$^{-1}$ \AA$^{-1}$ obtained from the F200W and F300M observations, respectively and this SED was well-represented by a blackbody model with 10680 K (see Figure \ref{F:SED2}) which is consistent with the previous value. This source does not have enough red color for a typical RSG while more likely to be a high-mass X-ray binary (HMXB). Moreover, its blackbody spectrum suggests that it is a highly probable donor candidate but, the possibility that the six NIR counterparts within the error radius of ULX-2 (see Figure \ref{F:multi}) could also be potential donor candidates should not be overlooked.\\

\section{CONCLUSIONS AND SUMMARY} \label{sec:5}

A challenge in determining the donor star in the optical emission of ULXs arises from the limited understanding of the flux contribution from the accretion disk, whether direct or irradiated. In addition, direct measurements of the masses of ULXs are quite difficult as they require dynamic mass measurements (radial velocity). However, constraints on their masses can be derived by studying their properties at multiple wavelengths. As seen in this study, the nature of ULXs can be constrained using optical and IR observations.\\

The significant improvement in sensitivity and resolution provided by {\it JWST} has led to a remarkable new perspective on the ambiguous nature of ULX donors, requiring a major reassessment of previous infrared imaging studies of ULX counterparts. We need to look for more counterparts with {\it JWST} data, and it is highly likely that the NIR counterparts reported in previous studies may have different characteristics than have been reported.\\

In this work, the first results of {\it JWST} NIR observations for the eight ULXs in NGC 1672 are presented. NIR counterparts of ULXs were searched by precise astrometry by using archival {\it NIRCam}, {\it HST}, and also {\ it GAIA} source catalog. At the same time, relative astrometry between {\it JWST} and {\it HST} was performed to assess NIR counterparts in the optical band. Evidence for warm dust or circumbinary disk was found for ULX-5 and ULX-8, and considerable clues were identified for the types of possible donor stars. Moreover, this work suggests that possible NIR counterparts of ULX-4 and ULX-8 which have isolated unique counterparts are ideal candidates for {\it JWST} Near-Infrared Spectrograph (NIRSpec) observations (0.1 arcsec/pixel). The main findings are summarized as follows:\\

\begin{itemize}

\item A unique NIR counterparts were identified for ULX-1, ULX-4, ULX-5 and ULX-8 and all unique counterparts were not determined mid-IR observations. The range absolute F200W ($\sim$ K-band)magnitudes of unique NIR counterparts is $-$ 8.5 < M$_{F200W}$ < $-$ 12.4. Among these ULXs, the optical counterpart was determined for only ULX-5. For ULX-2, ULX-3, ULX-6, and ULX-7, multiple NIR counterparts were determined. Except for ULX-3, at least one of them corresponds to a previously defined optical counterpart.\\

\item The SED for the counterparts of ULX-5 was well-fitted with a two-blackbody model with temperatures of 4300 (for optical) and 1600 K (for NIR excess). In this case, the optical emission may come from the possible late-type RSG.\\

\item We have reported the ULX-6 has a unique optical counterpart that can only be detected in the F814W filter. On the other hand, this study shows that this unique source actually is a blend of three sources that can be resolved in {\it NIRCam} images.\\

\item In the case of ULX-8 which does have not any optical counterpart, its NIR SED was well-represented by a blackbody with 1300 K. Similar to the NIR counterpart of ULX-5, the temperature obtained from the SED indicates that the ULX-8 system is in the warm dust or circumbinary disk. There is no clear evidence for the nature of the possible donor star of ULX-8 but, it is likely to be LMXB in dust heated by the compact object.\\

\item The NIR counterpart of ULX-4 could be AGN or star cluster due to its high K-band absolute magnitude (M$_{F200W}$ = -12 mag).\\

\item Finally, the SED of ULX-2 was re-plotted by adding two NIR data and it was well-fitted by a blackbody model with 10680 K which is consistent with the previous optical study.\\

\end{itemize}

\section*{Acknowledgements}
\noindent
This paper was supported by the Scientific and Technological Research Council of Turkey (TÜBİTAK) through project number 122C183. I thank the referee for the thorough and constructive comments that helped clarify and improve this paper. I would like to thank A. Akyuz for her valuable suggestions and support. I would like to extend my heartfelt appreciation to my wife Semiha Allak for her unwavering support, understanding and endless patience during the entire process of writing this paper. I would also like to thank my little son Atlas Allak who is the sweetest source of my motivation. Finally, many thanks to \cite{2023ApJ...944L..17L} for enabling the use of {\it JWST} images.

\section*{Data Availability}

The scientific results reported in this article are based on archival observations made by the James Webb Space Telescope and Hubble Space Telescope, and obtained from the data archive at the Space Telescope Science Institute\footnote{https://mast.stsci.edu/portal/Mashup/Clients/Mast/Portal.html}. This work has also made use of observations made with the {\it Chandra}\footnote{https://cda.harvard.edu/chaser/} X-ray Observatory.

\bibliographystyle{mnras}
\bibliography{ngc1672} % if your bibtex file is called 

\begin{thebibliography}{}
\makeatletter
\relax
\def\mn@urlcharsother{\let\do\@makeother \do\$\do\&\do\#\do\^\do\_\do\%\do\~}
\def\mn@doi{\begingroup\mn@urlcharsother \@ifnextchar [ {\mn@doi@} {\mn@doi@[]}}
\def\mn@doi@[#1]#2{\def\@tempa{#1}\ifx\@tempa\@empty \href {http://dx.doi.org/#2} {doi:#2}\else \href {http://dx.doi.org/#2} {#1}\fi \endgroup}
\def\mn@eprint#1#2{\mn@eprint@#1:#2::\@nil}
\def\mn@eprint@arXiv#1{\href {http://arxiv.org/abs/#1} {{\tt arXiv:#1}}}
\def\mn@eprint@dblp#1{\href {http://dblp.uni-trier.de/rec/bibtex/#1.xml} {dblp:#1}}
\def\mn@eprint@#1:#2:#3:#4\@nil{\def\@tempa {#1}\def\@tempb {#2}\def\@tempc {#3}\ifx \@tempc \@empty \let \@tempc \@tempb \let \@tempb \@tempa \fi \ifx \@tempb \@empty \def\@tempb {arXiv}\fi \@ifundefined {mn@eprint@\@tempb}{\@tempb:\@tempc}{\expandafter \expandafter \csname mn@eprint@\@tempb\endcsname \expandafter{\@tempc}}}

\bibitem[\protect\citeauthoryear{{Allak}}{{Allak}}{2022}]{2022MNRAS.517.3495A}
{Allak} S.,  2022, \mn@doi [\mnras] {10.1093/mnras/stac2887}, \href {https://ui.adsabs.harvard.edu/abs/2022MNRAS.517.3495A} {517, 3495}

\bibitem[\protect\citeauthoryear{{Allak} et~al.,}{{Allak} et~al.}{2022a}]{2022MNRAS.510.4355A}
{Allak} S.,  et~al., 2022a, \mn@doi [\mnras] {10.1093/mnras/stab3693}, \href {https://ui.adsabs.harvard.edu/abs/2022MNRAS.510.4355A} {510, 4355}

\bibitem[\protect\citeauthoryear{{Allak}, {Akyuz}, {Sonbas}  \& {Dhuga}}{{Allak} et~al.}{2022b}]{2022MNRAS.515.3632A}
{Allak} S.,  {Akyuz} A.,  {Sonbas} E.,   {Dhuga} K.~S.,  2022b, \mn@doi [\mnras] {10.1093/mnras/stac1992}, \href {https://ui.adsabs.harvard.edu/abs/2022MNRAS.515.3632A} {515, 3632}

\bibitem[\protect\citeauthoryear{{Ambrosi} \& {Zampieri}}{{Ambrosi} \& {Zampieri}}{2018}]{2018MNRAS.480.4918A}
{Ambrosi} E.,  {Zampieri} L.,  2018, \mn@doi [\mnras] {10.1093/mnras/sty2213}, \href {https://ui.adsabs.harvard.edu/abs/2018MNRAS.480.4918A} {480, 4918}

\bibitem[\protect\citeauthoryear{{Avdan}, {Akyuz}, {Acar}, {Akkaya Oralhan}, {Allak}  \& {Aksaker}}{{Avdan} et~al.}{2023}]{2023MNRAS.519.4826A}
{Avdan} S.,  {Akyuz} A.,  {Acar} S.,  {Akkaya Oralhan} I.,  {Allak} S.,   {Aksaker} N.,  2023, \mn@doi [\mnras] {10.1093/mnras/stac3811}, \href {https://ui.adsabs.harvard.edu/abs/2023MNRAS.519.4826A} {519, 4826}

\bibitem[\protect\citeauthoryear{{Bachetti} et~al.,}{{Bachetti} et~al.}{2014}]{2014Natur.514..202B}
{Bachetti} M.,  et~al., 2014, \mn@doi [\nat] {10.1038/nature13791}, \href {http://adsabs.harvard.edu/abs/2014Natur.514..202B} {514, 202}

\bibitem[\protect\citeauthoryear{{Barnes} et~al.,}{{Barnes} et~al.}{2022}]{2022A&A...662L...6B}
{Barnes} A.~T.,  et~al., 2022, \mn@doi [\aap] {10.1051/0004-6361/202243766}, 662, L6

\bibitem[\protect\citeauthoryear{{Bressan}, {Marigo}, {Girardi}, {Salasnich}, {Dal Cero}, {Rubele}  \& {Nanni}}{{Bressan} et~al.}{2012}]{2012MNRAS.427..127B}
{Bressan} A.,  {Marigo} P.,  {Girardi} L.,  {Salasnich} B.,  {Dal Cero} C.,  {Rubele} S.,   {Nanni} A.,  2012, \mn@doi [\mnras] {10.1111/j.1365-2966.2012.21948.x}, \href {https://ui.adsabs.harvard.edu/abs/2012MNRAS.427..127B} {427, 127}

\bibitem[\protect\citeauthoryear{{Brightman} et~al.,}{{Brightman} et~al.}{2018}]{2018NatAs...2..312B}
{Brightman} M.,  et~al., 2018, \mn@doi [Nature Astronomy] {10.1038/s41550-018-0391-6}, \href {https://ui.adsabs.harvard.edu/abs/2018NatAs...2..312B} {2, 312}

\bibitem[\protect\citeauthoryear{{Caballero-Garc{\'\i}a}, {Belloni}  \& {Zampieri}}{{Caballero-Garc{\'\i}a} et~al.}{2013}]{2013MNRAS.436.3262C}
{Caballero-Garc{\'\i}a} M.~D.,  {Belloni} T.,   {Zampieri} L.,  2013, \mn@doi [\mnras] {10.1093/mnras/stt1807}, \href {https://ui.adsabs.harvard.edu/abs/2013MNRAS.436.3262C} {436, 3262}

\bibitem[\protect\citeauthoryear{{Calamida} et~al.,}{{Calamida} et~al.}{2022}]{2022AJ....164...32C}
{Calamida} A.,  et~al., 2022, \mn@doi [\aj] {10.3847/1538-3881/ac73f0}, \href {https://ui.adsabs.harvard.edu/abs/2022AJ....164...32C} {164, 32}

\bibitem[\protect\citeauthoryear{{Carpano}, {Haberl}, {Maitra}  \& {Vasilopoulos}}{{Carpano} et~al.}{2018}]{2018MNRAS.476L..45C}
{Carpano} S.,  {Haberl} F.,  {Maitra} C.,   {Vasilopoulos} G.,  2018, \mn@doi [\mnras] {10.1093/mnrasl/sly030}, \href {https://ui.adsabs.harvard.edu/abs/2018MNRAS.476L..45C} {476, L45}

\bibitem[\protect\citeauthoryear{{Chandar}, {Whitmore}, {Dinino}, {Kennicutt}, {Chien}, {Schinnerer}  \& {Meidt}}{{Chandar} et~al.}{2016}]{2016ApJ...824...71C}
{Chandar} R.,  {Whitmore} B.~C.,  {Dinino} D.,  {Kennicutt} R.~C.,  {Chien} L.~H.,  {Schinnerer} E.,   {Meidt} S.,  2016, \mn@doi [\apj] {10.3847/0004-637X/824/2/71}, \href {https://ui.adsabs.harvard.edu/abs/2016ApJ...824...71C} {824, 71}

\bibitem[\protect\citeauthoryear{{Chen} \& {Podsiadlowski}}{{Chen} \& {Podsiadlowski}}{2019}]{2019ApJ...876L..11C}
{Chen} W.-C.,  {Podsiadlowski} P.,  2019, \mn@doi [\apjl] {10.3847/2041-8213/ab1b44}, \href {https://ui.adsabs.harvard.edu/abs/2019ApJ...876L..11C} {876, L11}

\bibitem[\protect\citeauthoryear{{Colbert} \& {Mushotzky}}{{Colbert} \& {Mushotzky}}{1999}]{1999ApJ...519...89C}
{Colbert} E. J.~M.,  {Mushotzky} R.~F.,  1999, \mn@doi [\apj] {10.1086/307356}, \href {https://ui.adsabs.harvard.edu/abs/1999ApJ...519...89C} {519, 89}

\bibitem[\protect\citeauthoryear{{Dudik}, {Berghea}, {Roberts}, {Gris{\'e}}, {Singh}, {Pagano}  \& {Winter}}{{Dudik} et~al.}{2016}]{2016ApJ...831...88D}
{Dudik} R.~P.,  {Berghea} C.~T.,  {Roberts} T.~P.,  {Gris{\'e}} F.,  {Singh} A.,  {Pagano} R.,   {Winter} L.~M.,  2016, \mn@doi [\apj] {10.3847/0004-637X/831/1/88}, \href {https://ui.adsabs.harvard.edu/abs/2016ApJ...831...88D} {831, 88}

\bibitem[\protect\citeauthoryear{{Eggleton}}{{Eggleton}}{1983}]{1983ApJ...268..368E}
{Eggleton} P.~P.,  1983, \mn@doi [\apj] {10.1086/160960}, \href {https://ui.adsabs.harvard.edu/abs/1983ApJ...268..368E} {268, 368}

\bibitem[\protect\citeauthoryear{{Elias}, {Frogel}  \& {Humphreys}}{{Elias} et~al.}{1985}]{1985ApJS...57...91E}
{Elias} J.~H.,  {Frogel} J.~A.,   {Humphreys} R.~M.,  1985, \mn@doi [\apjs] {10.1086/190997}, \href {https://ui.adsabs.harvard.edu/abs/1985ApJS...57...91E} {57, 91}

\bibitem[\protect\citeauthoryear{{Fabrika}, {Atapin}, {Vinokurov}  \& {Sholukhova}}{{Fabrika} et~al.}{2021}]{2021AstBu..76....6F}
{Fabrika} S.~N.,  {Atapin} K.~E.,  {Vinokurov} A.~S.,   {Sholukhova} O.~N.,  2021, \mn@doi [Astrophysical Bulletin] {10.1134/S1990341321010077}, \href {https://ui.adsabs.harvard.edu/abs/2021AstBu..76....6F} {76, 6}

\bibitem[\protect\citeauthoryear{{F{\"u}rst} et~al.,}{{F{\"u}rst} et~al.}{2016}]{2016ApJ...831L..14F}
{F{\"u}rst} F.,  et~al., 2016, \mn@doi [\apjl] {10.3847/2041-8205/831/2/L14}, \href {https://ui.adsabs.harvard.edu/abs/2016ApJ...831L..14F} {831, L14}

\bibitem[\protect\citeauthoryear{{Gris{\'e}}, {Kaaret}, {Corbel}, {Feng}, {Cseh}  \& {Tao}}{{Gris{\'e}} et~al.}{2012}]{2012ApJ...745..123G}
{Gris{\'e}} F.,  {Kaaret} P.,  {Corbel} S.,  {Feng} H.,  {Cseh} D.,   {Tao} L.,  2012, \mn@doi [\apj] {10.1088/0004-637X/745/2/123}, \href {https://ui.adsabs.harvard.edu/abs/2012ApJ...745..123G} {745, 123}

\bibitem[\protect\citeauthoryear{{Heida} et~al.,}{{Heida} et~al.}{2019}]{2019ApJ...883L..34H}
{Heida} M.,  et~al., 2019, \mn@doi [\apjl] {10.3847/2041-8213/ab4139}, \href {https://ui.adsabs.harvard.edu/abs/2019ApJ...883L..34H} {883, L34}

\bibitem[\protect\citeauthoryear{{Israel} et~al.,}{{Israel} et~al.}{2017a}]{2017Sci...355..817I}
{Israel} G.~L.,  et~al., 2017a, \mn@doi [Science] {10.1126/science.aai8635}, \href {https://ui.adsabs.harvard.edu/abs/2017Sci...355..817I} {355, 817}

\bibitem[\protect\citeauthoryear{{Israel} et~al.,}{{Israel} et~al.}{2017b}]{2017MNRAS.466L..48I}
{Israel} G.~L.,  et~al., 2017b, \mn@doi [\mnras] {10.1093/mnrasl/slw218}, \href {https://ui.adsabs.harvard.edu/abs/2017MNRAS.466L..48I} {466, L48}

\bibitem[\protect\citeauthoryear{{Jenkins} et~al.,}{{Jenkins} et~al.}{2011}]{2011ApJ...734...33J}
{Jenkins} L.~P.,  et~al., 2011, \mn@doi [\apj] {10.1088/0004-637X/734/1/33}, \href {https://ui.adsabs.harvard.edu/abs/2011ApJ...734...33J} {734, 33}

\bibitem[\protect\citeauthoryear{{Kaaret}, {Feng}  \& {Roberts}}{{Kaaret} et~al.}{2017}]{2017ARA&A..55..303K}
{Kaaret} P.,  {Feng} H.,   {Roberts} T.~P.,  2017, \mn@doi [\araa] {10.1146/annurev-astro-091916-055259}, \href {http://adsabs.harvard.edu/abs/2017ARA%26A..55..303K} {55, 303}

\bibitem[\protect\citeauthoryear{{King}}{{King}}{2009}]{2009MNRAS.393L..41K}
{King} A.~R.,  2009, \mn@doi [\mnras] {10.1111/j.1745-3933.2008.00594.x}, \href {http://adsabs.harvard.edu/abs/2009MNRAS.393L..41K} {393, L41}

\bibitem[\protect\citeauthoryear{{King}, {Lasota}  \& {Middleton}}{{King} et~al.}{2023}]{2023NewAR..9601672K}
{King} A.,  {Lasota} J.-P.,   {Middleton} M.,  2023, \mn@doi [\nar] {10.1016/j.newar.2022.101672}, \href {https://ui.adsabs.harvard.edu/abs/2023NewAR..9601672K} {96, 101672}

\bibitem[\protect\citeauthoryear{{Kong} et~al.,}{{Kong} et~al.}{2022}]{2022ApJ...933L...3K}
{Kong} L.-D.,  et~al., 2022, \mn@doi [\apjl] {10.3847/2041-8213/ac7711}, \href {https://ui.adsabs.harvard.edu/abs/2022ApJ...933L...3K} {933, L3}

\bibitem[\protect\citeauthoryear{{Lan{\c{c}}on}, {Hauschildt}, {Ladjal}  \& {Mouhcine}}{{Lan{\c{c}}on} et~al.}{2007}]{2007A&A...468..205L}
{Lan{\c{c}}on} A.,  {Hauschildt} P.~H.,  {Ladjal} D.,   {Mouhcine} M.,  2007, \mn@doi [\aap] {10.1051/0004-6361:20065824}, \href {https://ui.adsabs.harvard.edu/abs/2007A&A...468..205L} {468, 205}

\bibitem[\protect\citeauthoryear{{Lau} et~al.,}{{Lau} et~al.}{2019}]{2019ApJ...878...71L}
{Lau} R.~M.,  et~al., 2019, \mn@doi [\apj] {10.3847/1538-4357/ab1b1c}, \href {https://ui.adsabs.harvard.edu/abs/2019ApJ...878...71L} {878, 71}

\bibitem[\protect\citeauthoryear{{Lee} et~al.,}{{Lee} et~al.}{2023}]{2023ApJ...944L..17L}
{Lee} J.~C.,  et~al., 2023, \mn@doi [\apjl] {10.3847/2041-8213/acaaae}, \href {https://ui.adsabs.harvard.edu/abs/2023ApJ...944L..17L} {944, L17}

\bibitem[\protect\citeauthoryear{{Levesque}, {Massey}, {Olsen}, {Plez}, {Josselin}, {Maeder}  \& {Meynet}}{{Levesque} et~al.}{2005}]{2005ApJ...628..973L}
{Levesque} E.~M.,  {Massey} P.,  {Olsen} K.~A.~G.,  {Plez} B.,  {Josselin} E.,  {Maeder} A.,   {Meynet} G.,  2005, \mn@doi [\apj] {10.1086/430901}, \href {https://ui.adsabs.harvard.edu/abs/2005ApJ...628..973L} {628, 973}

\bibitem[\protect\citeauthoryear{{Liu}, {Bregman}  \& {McClintock}}{{Liu} et~al.}{2009}]{2009ApJ...690L..39L}
{Liu} J.,  {Bregman} J.~N.,   {McClintock} J.~E.,  2009, \mn@doi [\apjl] {10.1088/0004-637X/690/1/L39}, \href {https://ui.adsabs.harvard.edu/abs/2009ApJ...690L..39L} {690, L39}

\bibitem[\protect\citeauthoryear{{L{\'o}pez}, {Heida}, {Jonker}, {Torres}, {Roberts}, {Walton}, {Moon}  \& {Harrison}}{{L{\'o}pez} et~al.}{2020}]{2020MNRAS.497..917L}
{L{\'o}pez} K.~M.,  {Heida} M.,  {Jonker} P.~G.,  {Torres} M.~A.~P.,  {Roberts} T.~P.,  {Walton} D.~J.,  {Moon} D.~S.,   {Harrison} F.~A.,  2020, \mn@doi [\mnras] {10.1093/mnras/staa1920}, \href {https://ui.adsabs.harvard.edu/abs/2020MNRAS.497..917L} {497, 917}

\bibitem[\protect\citeauthoryear{{Middleton}, {Brightman}, {Pintore}, {Bachetti}, {Fabian}, {F{\"u}rst}  \& {Walton}}{{Middleton} et~al.}{2019}]{2019MNRAS.486....2M}
{Middleton} M.~J.,  {Brightman} M.,  {Pintore} F.,  {Bachetti} M.,  {Fabian} A.~C.,  {F{\"u}rst} F.,   {Walton} D.~J.,  2019, \mn@doi [\mnras] {10.1093/mnras/stz436}, \href {https://ui.adsabs.harvard.edu/abs/2019MNRAS.486....2M} {486, 2}

\bibitem[\protect\citeauthoryear{{Miller}, {Fabian}  \& {Miller}}{{Miller} et~al.}{2004}]{2004ApJ...614L.117M}
{Miller} J.~M.,  {Fabian} A.~C.,   {Miller} M.~C.,  2004, \mn@doi [\apjl] {10.1086/425316}, \href {https://ui.adsabs.harvard.edu/abs/2004ApJ...614L.117M} {614, L117}

\bibitem[\protect\citeauthoryear{{Morihana}, {Tsujimoto}, {Ebisawa}  \& {Gandhi}}{{Morihana} et~al.}{2022}]{2022PASJ...74..283M}
{Morihana} K.,  {Tsujimoto} M.,  {Ebisawa} K.,   {Gandhi} P.,  2022, \mn@doi [\pasj] {10.1093/pasj/psab124}, \href {https://ui.adsabs.harvard.edu/abs/2022PASJ...74..283M} {74, 283}

\bibitem[\protect\citeauthoryear{{Motch}, {Pakull}, {Soria}, {Gris{\'e}}  \& {Pietrzy{\'n}ski}}{{Motch} et~al.}{2014}]{2014Natur.514..198M}
{Motch} C.,  {Pakull} M.~W.,  {Soria} R.,  {Gris{\'e}} F.,   {Pietrzy{\'n}ski} G.,  2014, \mn@doi [\nat] {10.1038/nature13730}, \href {https://ui.adsabs.harvard.edu/abs/2014Natur.514..198M} {514, 198}

\bibitem[\protect\citeauthoryear{{Nelson}, {Rappaport}  \& {Joss}}{{Nelson} et~al.}{1986}]{1986ApJ...311..226N}
{Nelson} L.~A.,  {Rappaport} S.~A.,   {Joss} P.~C.,  1986, \mn@doi [\apj] {10.1086/164767}, \href {https://ui.adsabs.harvard.edu/abs/1986ApJ...311..226N} {311, 226}

\bibitem[\protect\citeauthoryear{{Rappaport} \& {Joss}}{{Rappaport} \& {Joss}}{1984}]{1984ApJ...283..232R}
{Rappaport} S.,  {Joss} P.~C.,  1984, \mn@doi [\apj] {10.1086/162298}, \href {https://ui.adsabs.harvard.edu/abs/1984ApJ...283..232R} {283, 232}

\bibitem[\protect\citeauthoryear{{Roberts}}{{Roberts}}{2007}]{2007Ap&SS.311..203R}
{Roberts} T.~P.,  2007, \mn@doi [\apss] {10.1007/s10509-007-9547-z}, \href {https://ui.adsabs.harvard.edu/abs/2007Ap&SS.311..203R} {311, 203}

\bibitem[\protect\citeauthoryear{{Rodr{\'\i}guez Castillo} et~al.,}{{Rodr{\'\i}guez Castillo} et~al.}{2020}]{2020ApJ...895...60R}
{Rodr{\'\i}guez Castillo} G.~A.,  et~al., 2020, \mn@doi [\apj] {10.3847/1538-4357/ab8a44}, \href {https://ui.adsabs.harvard.edu/abs/2020ApJ...895...60R} {895, 60}

\bibitem[\protect\citeauthoryear{{Rodr{\'\i}guez} et~al.,}{{Rodr{\'\i}guez} et~al.}{2023}]{2023ApJ...944L..26R}
{Rodr{\'\i}guez} M.~J.,  et~al., 2023, \mn@doi [\apjl] {10.3847/2041-8213/aca653}, \href {https://ui.adsabs.harvard.edu/abs/2023ApJ...944L..26R} {944, L26}

\bibitem[\protect\citeauthoryear{{Sathyaprakash} et~al.,}{{Sathyaprakash} et~al.}{2019}]{2019MNRAS.488L..35S}
{Sathyaprakash} R.,  et~al., 2019, \mn@doi [\mnras] {10.1093/mnrasl/slz086}, \href {https://ui.adsabs.harvard.edu/abs/2019MNRAS.488L..35S} {488, L35}

\bibitem[\protect\citeauthoryear{{Sathyaprakash} et~al.,}{{Sathyaprakash} et~al.}{2022}]{2022MNRAS.511.5346S}
{Sathyaprakash} R.,  et~al., 2022, \mn@doi [\mnras] {10.1093/mnras/stac402}, \href {https://ui.adsabs.harvard.edu/abs/2022MNRAS.511.5346S} {511, 5346}

\bibitem[\protect\citeauthoryear{{Soria}, {Hakala}, {Hau}, {Gladstone}  \& {Kong}}{{Soria} et~al.}{2012}]{2012MNRAS.420.3599S}
{Soria} R.,  {Hakala} P.~J.,  {Hau} G. K.~T.,  {Gladstone} J.~C.,   {Kong} A. K.~H.,  2012, \mn@doi [\mnras] {10.1111/j.1365-2966.2011.20281.x}, \href {https://ui.adsabs.harvard.edu/abs/2012MNRAS.420.3599S} {420, 3599}

\bibitem[\protect\citeauthoryear{{Stetson}}{{Stetson}}{1987}]{1987PASP...99..191S}
{Stetson} P.~B.,  1987, \mn@doi [\pasp] {10.1086/131977}, \href {https://ui.adsabs.harvard.edu/abs/1987PASP...99..191S} {99, 191}

\bibitem[\protect\citeauthoryear{{Sutton}, {Roberts}, {Walton}, {Gladstone}  \& {Scott}}{{Sutton} et~al.}{2012}]{2012MNRAS.423.1154S}
{Sutton} A.~D.,  {Roberts} T.~P.,  {Walton} D.~J.,  {Gladstone} J.~C.,   {Scott} A.~E.,  2012, \mn@doi [\mnras] {10.1111/j.1365-2966.2012.20944.x}, \href {https://ui.adsabs.harvard.edu/abs/2012MNRAS.423.1154S} {423, 1154}

\bibitem[\protect\citeauthoryear{{Sutton}, {Done}  \& {Roberts}}{{Sutton} et~al.}{2014}]{2014MNRAS.444.2415S}
{Sutton} A.~D.,  {Done} C.,   {Roberts} T.~P.,  2014, \mn@doi [\mnras] {10.1093/mnras/stu1597}, \href {https://ui.adsabs.harvard.edu/abs/2014MNRAS.444.2415S} {444, 2415}

\bibitem[\protect\citeauthoryear{{Tao}, {Feng}, {Gris{\'e}}  \& {Kaaret}}{{Tao} et~al.}{2011}]{2011ApJ...737...81T}
{Tao} L.,  {Feng} H.,  {Gris{\'e}} F.,   {Kaaret} P.,  2011, \mn@doi [\apj] {10.1088/0004-637X/737/2/81}, \href {http://adsabs.harvard.edu/abs/2011ApJ...737...81T} {737, 81}

\bibitem[\protect\citeauthoryear{{Tao}, {Kaaret}, {Feng}  \& {Gris{\'e}}}{{Tao} et~al.}{2012}]{2012ApJ...750..110T}
{Tao} L.,  {Kaaret} P.,  {Feng} H.,   {Gris{\'e}} F.,  2012, \mn@doi [\apj] {10.1088/0004-637X/750/2/110}, \href {https://ui.adsabs.harvard.edu/abs/2012ApJ...750..110T} {750, 110}

\bibitem[\protect\citeauthoryear{{Walton} et~al.,}{{Walton} et~al.}{2018}]{2018ApJ...857L...3W}
{Walton} D.~J.,  et~al., 2018, \mn@doi [\apjl] {10.3847/2041-8213/aabadc}, \href {https://ui.adsabs.harvard.edu/abs/2018ApJ...857L...3W} {857, L3}

\bibitem[\protect\citeauthoryear{{Yang} \& {Jiang}}{{Yang} \& {Jiang}}{2011}]{2011ApJ...727...53Y}
{Yang} M.,  {Jiang} B.~W.,  2011, \mn@doi [\apj] {10.1088/0004-637X/727/1/53}, \href {https://ui.adsabs.harvard.edu/abs/2011ApJ...727...53Y} {727, 53}

\bibitem[\protect\citeauthoryear{{Yang} \& {Jiang}}{{Yang} \& {Jiang}}{2012}]{2012ApJ...754...35Y}
{Yang} M.,  {Jiang} B.~W.,  2012, \mn@doi [\apj] {10.1088/0004-637X/754/1/35}, \href {https://ui.adsabs.harvard.edu/abs/2012ApJ...754...35Y} {754, 35}

\bibitem[\protect\citeauthoryear{{Yao} \& {Feng}}{{Yao} \& {Feng}}{2019}]{2019ApJ...884L...3Y}
{Yao} Y.,  {Feng} H.,  2019, \mn@doi [\apjl] {10.3847/2041-8213/ab44c7}, \href {https://ui.adsabs.harvard.edu/abs/2019ApJ...884L...3Y} {884, L3}

\bibitem[\protect\citeauthoryear{{de Naray}, {Brandt}, {Halpern}  \& {Iwasawa}}{{de Naray} et~al.}{2000}]{2000AJ....119..612D}
{de Naray} P.~J.,  {Brandt} W.~N.,  {Halpern} J.~P.,   {Iwasawa} K.,  2000, \mn@doi [\aj] {10.1086/301220}, \href {https://ui.adsabs.harvard.edu/abs/2000AJ....119..612D} {119, 612}

\makeatother
\end{thebibliography}

\bsp	% typesetting comment
\label{lastpage}
\end{document}